%% file: newv2.tex
\documentclass[12pt]{iopart}

\usepackage{graphicx, amssymb, amsfonts, algorithm, algorithmic}
\usepackage{cite}
\newcommand{\hH}{\hat{H}}
\newcommand{\hs}{\hat{\sigma}}

\newcommand{\ket}[1]{| #1 \rangle}

\newcommand{\bra}[1]{\langle #1 |}
\newcommand{\overlap}[2]{\langle #1|#2\rangle}
\newcommand{\expectation}[2]{\langle #1|#2|#1\rangle}
\newcommand{\mean}[1]{\langle #1\rangle}

\begin{document}

\title{A quantum algorithm to count weighted ground states of classical spin Hamiltonians}

\author{Bhuvanesh Sundar$^{1,2}$, Roger Paredes$^3$, David T. Damanik$^4$, Leonardo Due{\~n}as-Osorio$^3$ and Kaden R. A. Hazzard$^{1,2}$}
\address{$^1$Department of Physics and Astronomy, Rice University, Houston, Texas 77005, USA}
\address{$^2$Rice Center for Quantum Materials, Rice University, Houston, Texas 77005, USA}
\address{$^3$Department of Civil and Environmental Engineering, Rice University, Houston, Texas 77005, USA}
\address{$^4$Department of Mathematics, Rice University, Houston, Texas 77005, USA}
\eads{\mailto{Bhuvanesh.Sundar@rice.edu}, \mailto{roger.paredes@rice.edu}, \mailto{David.T.Damanik@rice.edu}, \mailto{leonardo.duenas-osorio@rice.edu}, \mailto{kaden@rice.edu}}

\begin{abstract}
Ground state counting plays an important role in several applications in science and engineering, from estimating residual entropy in physical systems, to bounding engineering reliability and solving combinatorial counting problems. While quantum algorithms such as adiabatic quantum optimization (AQO) and quantum approximate optimization (QAOA) can minimize Hamiltonians, they are inadequate for counting ground states. We modify AQO and QAOA to count the ground states of arbitrary classical spin Hamiltonians, including counting ground states with arbitrary nonnegative weights attached to them. As a concrete example, we show how our method can be used to count the weighted fraction of edge covers on graphs, with user-specified confidence on the relative error of the weighted count, in the asymptotic limit of large graphs. We find the asymptotic computational time complexity of our algorithms, via analytical predictions for AQO and numerical calculations for QAOA, and compare with the classical optimal Monte Carlo algorithm (OMCS), as well as a modified Grover's algorithm. We show that for large problem instances with small weights on the ground states, AQO does not have a quantum speedup over OMCS for a fixed error and confidence, but QAOA has a sub-quadratic speedup on a broad class of numerically simulated problems. Our work is an important step in approaching general ground-state counting problems beyond those that can be solved with Grover's algorithm. It offers algorithms that can employ noisy intermediate-scale quantum devices for solving ground state counting problems on small instances, which can help in identifying more problem classes with quantum speedups.
\end{abstract}

\maketitle

\section{Introduction}
\input{intro}

\begin{figure}[t]\centering
\includegraphics[width=0.8\columnwidth]{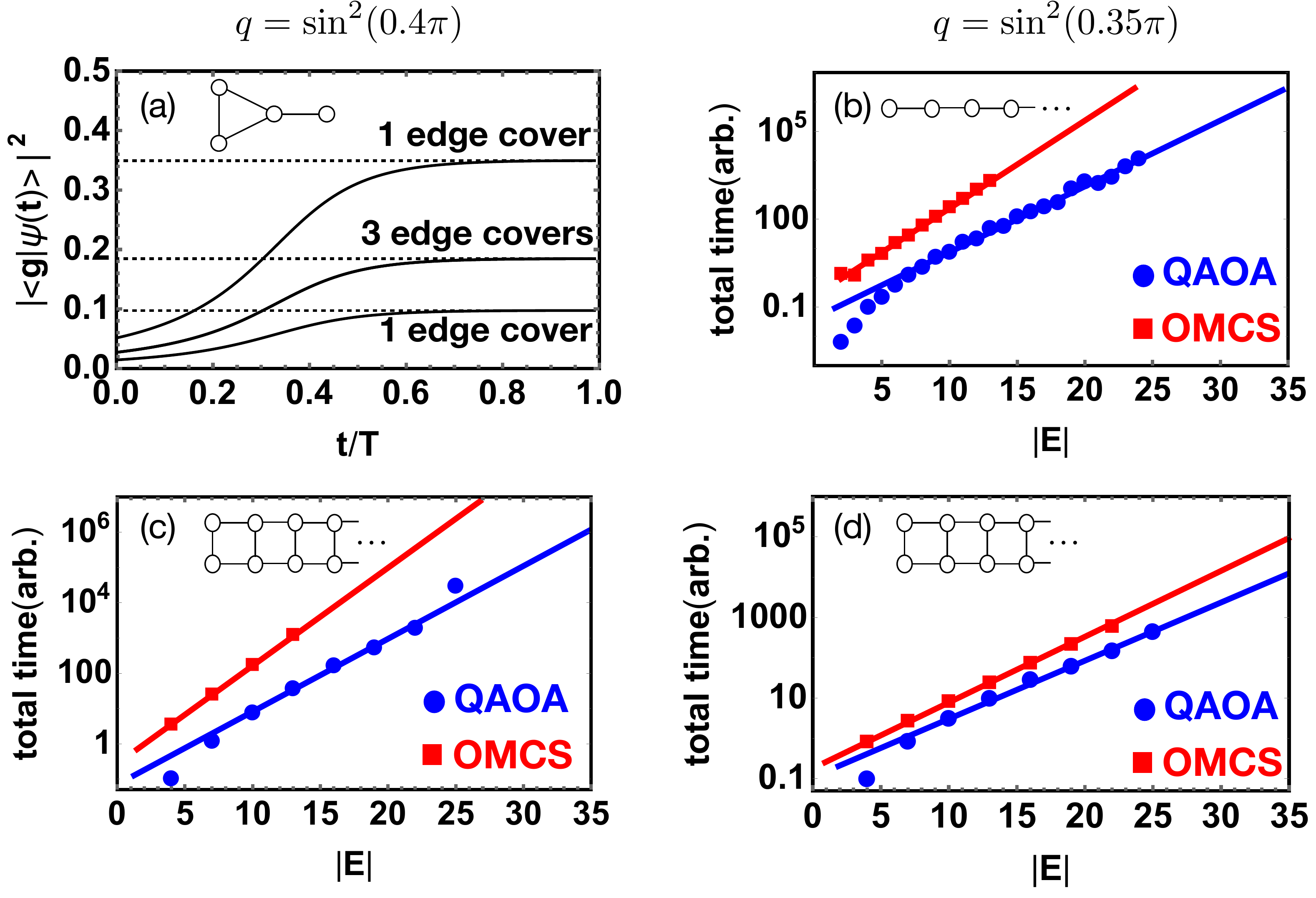}
\caption{(Color online) Importance-sampling of ground states in AQO, and the scaling of total computational time for QAOA and OMCS. (a) Demonstration that the instantaneous wave function in AQO importance-samples the ground states of $\hH_z$ in Eq.~(\ref{eqn: Hz}). Solid lines plot $|\overlap{g}{\psi(t)}|^2$, for all the ground states of $\hH_z$ encoding the edge cover problem for the paw graph---shown in the panel and described in Fig.~\ref{fig: edge covers} and Sec.~\ref{subsec: edge covers}. Dashed lines plot $|\overlap{g}{\psi(t)}|^2/\langle\psi(t)|\hat{P}_\mathcal{G}|\psi(t)\rangle$, where $\hat{P}_\mathcal{G}$ is the projection operator onto the ground state space $\mathcal{G}$ of $\hH_z$. The paw graph has five degenerate ground states for $\hH_z$, with one state having weight $w=q^2(1-q)^2$, three with $w=q(1-q)^3$ (whose solid curves as well as dashed curves overlap), and one with $w=(1-q)^4$. The overlapping solid curves and the flat dashed curves all illustrate that the wave function importance-samples the ground states at all times, i.e., $|\overlap{g}{\psi(t)}|^2 \propto w(g)$. 
(b)-(d) Scaling of the total computational time taken by QAOA (blue circles) vs. OMCS (red squares) to estimate the weighted count $\mathcal{P}$ of edge covers for the graphs shown in the respective panels and $q$ above the panels, with probability $1-\delta=0.95$ of having the relative error less than $\epsilon=0.05$. The ``total time'' plotted for OMCS is the physical CPU time in seconds, while the ``time'' for QAOA is the total number of one-qubit gates and CNOTs in repeated iterations of the optimal QAOA circuit found by greedy variational optimization described in Sec.~\ref{subsec: QAOA}, multiplied by a constant factor to lie on the same scale as OMCS. In all these cases, QAOA is asymptotically faster than OMCS, as seen by extrapolating the results to large $|E|$ (solid lines). The plots do not include the time $T_{\rm\alpha\beta\ search}$ to find the variational parameters in QAOA.}
\label{fig: main results}
\end{figure}

\begin{table}\centering \begin{tabular}{|c|c|c|c|}
\hline
Algorithm & $\begin{array}{c}{\rm Number\ of\ steps\ to}\\ {\rm reach\ ground\ states}\end{array}$ & $\begin{array}{c}{\rm Number\ of\ measurements}\\ {\rm to\ estimate\ }\mathcal{P} \end{array}$ & Source\\
\hline
$\begin{array}{c}{\rm Classical\ Optimal}\\{\rm Monte\ Carlo (OMCS)}\end{array}$ & $\frac{1}{\mathcal{P}}$ & $\frac{|\ln(\delta)|}{\epsilon^2}$ & \cite{karp1983monte}\\
$\begin{array}{c}{\rm Adiabatic\ quantum}\\{\rm optimization\ (AQO)}\end{array}$ & $T_{\rm AQO}/\rmd t \sim \frac{1}{\eta\mathcal{P}}$ &  $T_{\rm count} = \frac{\sqrt{|\ln(\delta)|}}{\epsilon(1-\eta^2)}\sqrt{\frac{ \mathcal{P}^2 }{ \mathcal{P}_2 }}$ & this paper\\
$\begin{array}{c}{\rm Quantum\ approximate}\\{\rm optimization\ (QAOA)}\end{array}$ & $ T_{\rm QAOA} \sim \frac{\sin^{-1}\sqrt{1-\eta^2}}{\sqrt{\mathcal{P}}}$ & $T_{\rm count} = \frac{\sqrt{|\ln(\delta)|}}{\epsilon(1-\eta^2)}\sqrt{\frac{ \mathcal{P}^2 }{ \mathcal{P}_2 }}$ & this paper\\
Grover's algorithm & $T_{\rm Grover} \sim \frac{\sin^{-1}\sqrt{1-\eta^2}} {2\sqrt{\mathcal{P}}}$ & $ T_{\rm count} = \frac{\sqrt{|\ln(\delta)|}}{\epsilon(1-\eta^2)}\sqrt{\frac{ \mathcal{P}^2 }{ \mathcal{P}_2 }}$ & \cite{grover1998quantum} + this paper\\
\hline
\end{tabular}
\caption{Scaling of the number of operations required by different algorithms to estimate the weighted ground state count $\mathcal{P}$ for a classical Hamiltonian $\hH_z$, with maximum relative error $\epsilon$ and confidence $1-\delta$ in the limit $\epsilon,\delta\rightarrow0$ [see Eq.~(\ref{eqn: confidence})]. Second column: Number of random samples drawn to find a ground state in OMCS, and the number of calls to $\hH_x$ and $\hH_z$ in AQO and QAOA, and oracle calls in Grover's algorithm, to reach ground state occupation $\expectation{\psi(T)}{\hat{P}_\mathcal{G}} = 1-\eta^2$. Third column: Number of measurements made until the statistical analysis yields $\mathcal{P}$ with relative error $\epsilon$ and confidence $1-\delta$. For OMCS, this column refers to the number of ground states measured. $\mathcal{P}_2$ is the sum of squares of the ground state weights [see Eq.~(\ref{eqn: moments})]. The total time in all these algorithms scales as the product of the second column, the third column, and the time required to implement one step of the second column (e.g., draw one random sample in OMCS and verify if its a ground state). The scaling quoted for QAOA is found numerically, and does not include the time $T_{\rm\alpha\beta\ search}$ to find the variational parameters in QAOA. The total computational time has additional factors not listed here, many of them varying polynomially with the number of qubits and discussed in Sec.~\ref{sec: methods}.}
\label{tab}
\end{table}

\section{Problem: Counting ground states of a classical Hamiltonian}\label{sec: setup}

\input{generalproblem}

\input{edgecovers}
\begin{figure}[t]\centering
\includegraphics[width=0.75\columnwidth]{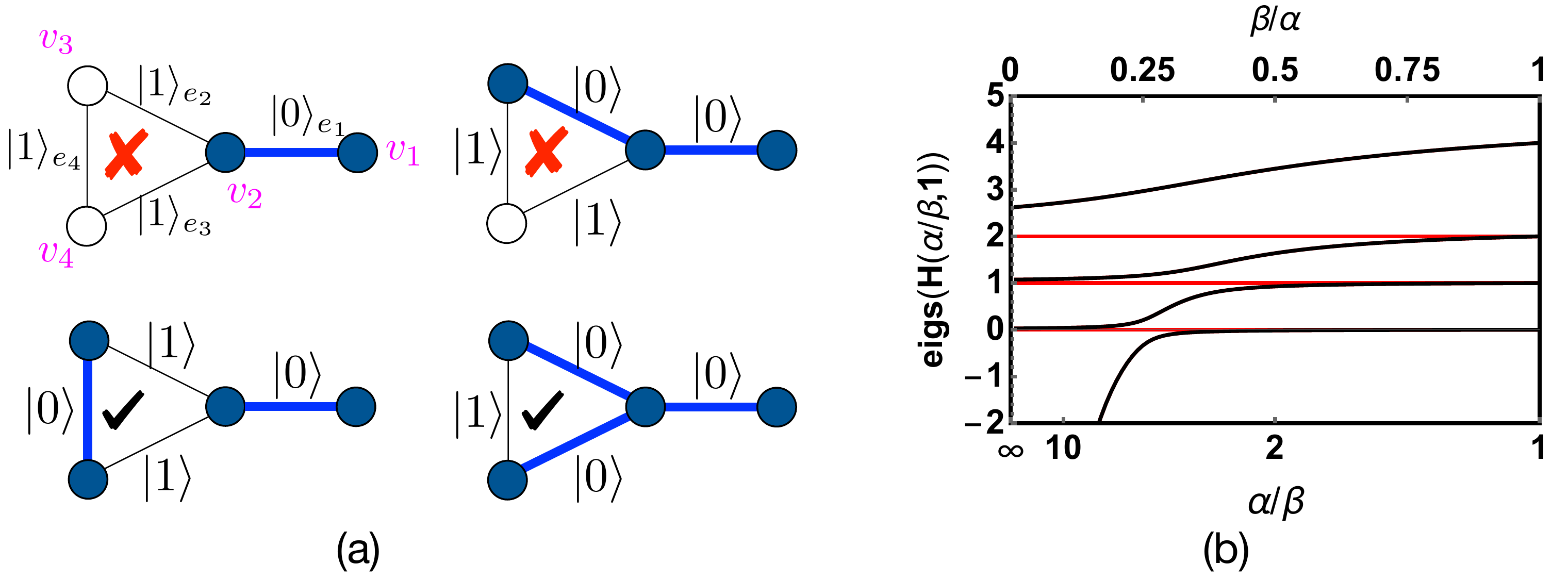}
\caption{Examples of edge covers and non-edge-covers, and the spectrum of $\hH(\alpha/\beta,1) = (\alpha/\beta) \hH_x+\hH_z$ , for the paw graph. The Hamiltonians $\hH_z$ and $\hH_x$ are defined in Eqs.~(\ref{eqn: Hz}) and (\ref{eqn: Hx}). (a) The top two panels show examples where the set of thick blue links (denoted $E'$ in the text) are not edge covers, and bottom panels show examples that are edge covers. For illustration, vertices incident to links in $E'$ are shaded blue; $E'$ is an edge cover if all vertices are shaded. Out of the sixteen subsets on this graph, five are edge covers and ground states of Eq.~(\ref{eqn: Hz}), with thick links mapped to $\ket{0}$ and thin links mapped to $\ket{1}$. (b) Spectrum of $\hH(\alpha/\beta,1)$ for the paw graph. The flat red lines are the energies for the antisymmetric eigenstates of $\hH$, and solid black lines are the energies for the symmetric eigenstates at $q=\sin^2(0.4\pi)$. The minimum value of the difference between the two lowest black lines determines the evolution time $T_{\rm AQO}$ in AQO [Eqs.~(\ref{eqn: adia thm}) and~(\ref{eqn: TAQO full})].}
\label{fig: edge covers}
\end{figure}

\section{Methods: Algorithms for importance-sampling and counting}\label{sec: methods}
\input{methods_v2}

\subsection{Grover's algorithm with importance-sampling.}\label{subsec: Grover}

\input{grover}

\subsection{AQO with importance-sampling.}\label{subsec: QA}

\begin{figure}[t]\centering
\includegraphics[width=0.8\columnwidth]{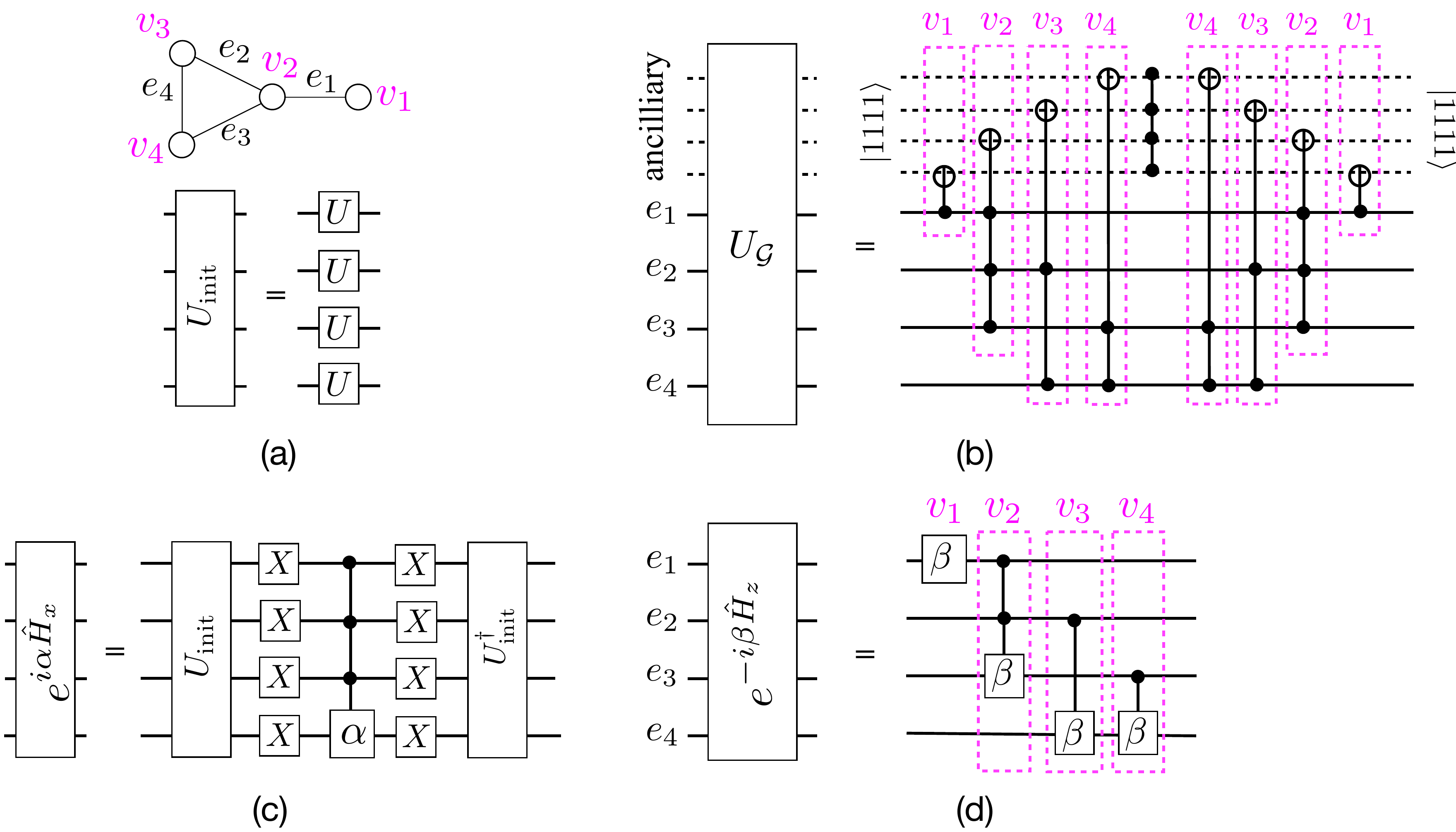}
\caption{Quantum circuits to prepare the initial quantum state and implement one discrete quantum evolution step, for the paw graph shown in (a). (a) Circuit to prepare the initial state $\ket{\psi(0)}$ [in Eq.~(\ref{eqn: Grover})] for the weight function in Eq.~(\ref{eqn: weights}), when the input state is $\ket{00\cdots}$. $U=\exp(-i\sigma^y \sin^{-1}\sqrt{q})$ is a one-qubit unitary operator. Preparing $\ket{\psi(0)}$ for more general weight functions is non-trivial. (b) Implementation of the Grover oracle $U_\mathcal{G}$ for $\hH_z$ in Eq.~(\ref{eqn: Hz}). The circuit has four ancillary bits (dashed lines), one each to verify the local constraint satisfaction for the corresponding node labeled in pink. Circuits with fewer or no ancillary bits may be possible. (c) Implementation of $\exp(i\alpha\hH_x)$, where the many-qubit gate is the controlled-phase gate with phase $\exp(-i\alpha)$. This circuit also implements the Grover diffusion operator $U_0$ [in Eq.~(\ref{eqn: Grover})] when $\alpha=\pi$. (d) Implementation of $\exp(-i\beta\hH_z)$, where the single-qubit gate is $e^{-i\beta}\ket{1}\bra{1}$, and the multi-qubit gates are controlled-phase gates with phase $\exp(-i\beta)$.}
\label{fig: qtm cct}
\end{figure}

\begin{figure}[t]\centering
\includegraphics[width=\columnwidth]{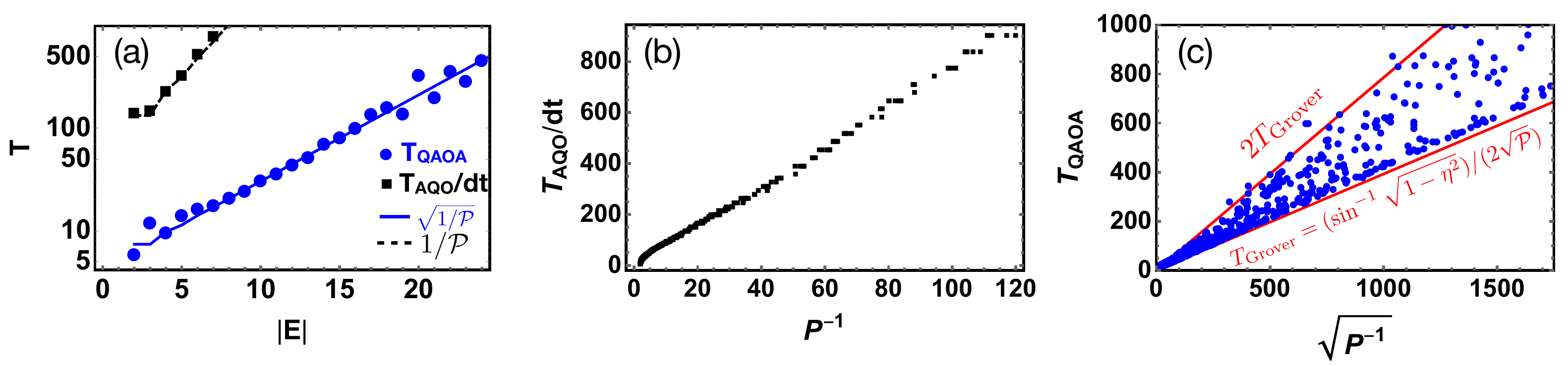}
\caption{(Color online)
The number of discrete AQO steps $T_{\rm AQO}/\rmd t$ and the number of QAOA steps $T_{\rm QAOA}$ in a numerical simulation of these algorithms, until the system reaches desired ground state occupation $\expectation{\psi(T)} { \hat{P}_\mathcal{G} } = 1-\eta^2$. 
(a) $T_{\rm AQO}/\rmd t$ (black squares) and $T_{\rm QAOA}$ (blue circles) required to reach $1-\eta^2=0.8$, for $q=\sin^2(0.3\pi)$ on linear graphs. These two curves scale the same way with the system size as $1/\mathcal{P}$ (black dashed line) and $1/\sqrt{\mathcal{P}}$ (blue solid line) respectively up to overall polynomial prefactors. 
For the class of graphs and $q$ considered here, $1/\mathcal{P} \sim 1.47^{|E|}$ [see also Eq.~(\ref{eqn: analyticalP}) for a closed form]. 
(b) $T_{\rm AQO}/\rmd t$ required to reach $1-\eta^2=0.5$ for an ensemble of random graphs with mean vertex degrees $1.25$ and $2.5$, $|E|$ ranging from $5$ to $25$, and $q$ varying from $0$ to $1$. We chose $\rmd t=0.1$. $T_{\rm AQO}$ scales as $1/\mathcal{P}$, consistent with the analytical prediction in Eq.~(\ref{eqn: TAQO}). 
(c) $T_{\rm QAOA}$ required to reach $1-\eta^2=0.5$ for the same ensemble of graphs and parameters as (b). For this ensemble, $T_{\rm QAOA}$ mostly lies between $(\sin^{-1}\sqrt{1-\eta^2})/(2\sqrt{\mathcal{P}})$ and $(\sin^{-1}\sqrt{1-\eta^2})/\sqrt{\mathcal{P}}$. Notably, the number of Grover iterations required to reach the same ground state occupation is $T_{\rm Grover}=(\sin^{-1}\sqrt{1-\eta^2})/(2\sqrt{\mathcal{P}})$. Only points with $T_{\rm QAOA}<1000$ and $T_{\rm AQO}/\rmd t<1000$ are shown.
}
\label{fig: AQO and qaoa times}
\end{figure}

\input{aqo_v2}

\subsection{QAOA with importance-sampling.}\label{subsec: QAOA}

\begin{figure}[t]\centering
\includegraphics[width=0.85\columnwidth]{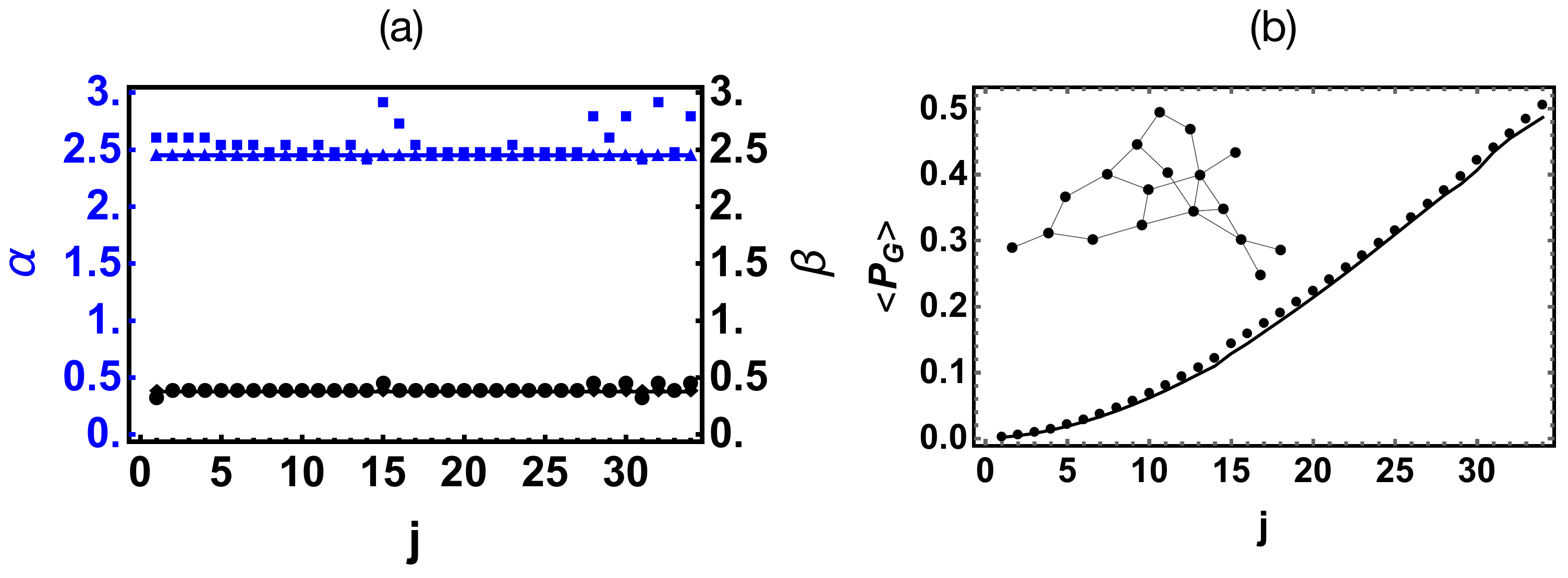}
\caption{(Color online) (a) The variationally optimized $\alpha_j$ and $\beta_j$ in QAOA, and (b) the instantaneous projection $\expectation{\psi(j)}{ \hat{P}_\mathcal{G} }$, for the random graph shown in the inset in (b). The points show the numerical results for $\alpha_j^{\rm opt}$ and $\beta_j^{\rm opt}$ obtained from greedy optimization, while the lines show the results when $\alpha^{\rm opt}=0.78\pi$ and $\beta^{\rm opt}=0.12\pi$ are variationally obtained constants. Variationally finding $\alpha_j^{\rm opt}$ and $\beta_j^{\rm opt}$ with the greedy method for $j=1, \cdots, T_{\rm QAOA}$ takes time $T_{\rm \alpha\beta\ search}^{\rm greedy} \sim T_{\rm QAOA}^2$, while finding constant $\alpha^{\rm opt}$ and $\beta^{\rm opt}$ might require only $T_{\rm \alpha\beta\ search}^{\rm constant} \sim O(1)$.}
\label{fig: qaoa with constant angles}
\end{figure}

\input{qaoa}

\input{counting}

\input{results_v2}

\input{summary}

\section*{Acknowledgments}
This material is based upon work supported with funds from the Welch Foundation  Grant no.  C-1872 and from NSF Grant No.~PHY-1848304. We thank Rice University for its Creative Ventures Funds Program and the InterDisciplinary Excellence Awards (IDEA). We acknowledge the use of IBM Q Experience~\cite{qiskit} for this work. We thank Moshe Vardi and Anastasios Kyrillidis, professors in computer science at Rice University, for several useful conversations.

\section*{References}
\bibliography{BibFile}

\input{appendix}
\input{algo}

\end{document}

%% file: intro.tex
Counting ground states of classical spin Hamiltonians (or equivalently, global minima of functions of binary variables) is a computationally difficult problem that finds wide applications in science and engineering. Many problems of practical importance, such as probabilistic reasoning and Bayesian inference~\cite{li2006performing,chavira2008probabilistic,sang2005performing,littman2001stochastic,davies2007using,domshlak2006fast}, determining the reliability of graph flows for energy, information, and mechanical structures~\cite{paredes2018network,khazaei2018smart}, membership filters~\cite{weaver2014satisfiability,douglass2015constructing,azinovic2017assessment,biere2009handbook}, and performing data-driven diagnosis~\cite{kumar2002model}, rely on counting minima of cost functions which encode relevant constraints. In physical systems, ground state degeneracy arises from geometric frustration~\cite{moessner2006geometrical}, glassy physics~\cite{binder1986spin,castellani2005spin}, and novel ordering~\cite{balents2010spin,sadoc2006geometrical}.

Adiabatic quantum optimization (AQO)~\cite{farhi2000quantum,de2011introduction} and, more recently, a hybrid classical-quantum variational algorithm called quantum approximate optimization (QAOA)~\cite{farhi2014quantum2,farhi2014quantum1}, are two algorithms widely used~\cite{venegas2018cross,das2005quantum,santoro2006optimization,das2008quantum,albash2018adiabatic,finnila1994quantum,kadowaki1998quantum,santoro2002theory,cohen2015quantum,albash2018demonstration,muthukrishnan2016tunneling,hen2010probing,denchev2016computational,mandra2016strengths,farhi2016quantum,zhou2018quantum,crooks2018performance,harrow2019low,guerreschi2017practical,gilyen2019optimizing,hadfield2019quantum,peruzzo2014variational,kokail2018self,pagano2019quantum,moll2018quantum,wecker2016training,wecker2015progress,verdon2017quantum,verdon2019quantum,hastings2019classical,wang2019xy,wang2018quantum,morales2018variational,mbeng2019quantum,parrish2019jacobi,niu2019optimizing,campbell2019applying,akshay2019reachability,shaydulin2019multistart,bapat2018bang,guerreschi2019qaoa} to minimize spin Hamiltonians, including several that solve hard optimization problems in science and engineering. Excitingly, QAOA has the potential to be implemented on current noisy intermediate-scale quantum (NISQ) devices~\cite{kokail2018self,pagano2019quantum,moll2018quantum}.

However, despite their promise of finding a ground state of these Hamiltonians, AQO and QAOA are inefficient for counting their ground states~\cite{boixo2013experimental,konz2018uncertain,matsuda2009quantum,king2016degeneracy,mandra2017exponentially,zhang2017advantages,katzgraber2018viewing,matsuda2009ground}, when implemented in their usual form with a transverse field as the mixing Hamiltonian. This is because they result in a final wave function with a small or zero weight on a significant number of the classical ground states. Adaptations of AQO and QAOA that solve counting problems must ensure that the amplitudes of the final wave function in these algorithms sample all the classical ground states with sufficient probability.

In this work, we modify AQO and QAOA to count ground states of arbitrary classical spin Hamiltonians. Our work is inspired by ideas in Refs.~\cite{matsuda2009ground,hen2014fast,van2001powerful,roland2002quantum} to fairly sample ground states, which are in turn inspired by Grover's algorithm~\cite{grover1997quantum}. Additionally, we extend these algorithms to count ground states with arbitrary weights attached to them, by designing the algorithms such that the final wave function importance-samples the ground states with probabilities given by their weights.

We demonstrate our algorithms by applying them to count weighted edge covers on graphs. This is directly related to calculating the edge cover polynomial of a graph~\cite{akbari2013edge}, and has applications in reliability engineering~\cite{paredes2018network,duenas2018quantum}. We compare the performance of our algorithms versus optimal Monte Carlo simulation (OMCS)~\cite{karp1983monte,dagum2000optimal}, which is a widely used classical method to numerically simulate engineering problems, that \textit{a priori} provides confidence on the relative error of the expectations of random variables with minimal assumptions.

The main results presented in this article, shown in Fig.~\ref{fig: main results} and in Table~\ref{tab} with relevant notations defined in Secs.~\ref{sec: setup} and~\ref{sec: methods}, are as follows: (1) We show that the wave function in our algorithms,  at any time during their execution, has amplitudes that importance-sample the ground states of a classical Hamiltonian, and (2) We analyze the asymptotic scaling of the time required by these algorithms to estimate the weighted count of the ground states, analytically in the case of AQO with an arbitrary classical Hamiltonian, and numerically in the case of QAOA to count edge covers. We find that (a) AQO with a linear schedule is slower than OMCS for a given relative error and confidence, but (b) QAOA can have a speedup over classical OMCS when the total weight on the ground states is small. The speedup is sub-quadratic, and assumes that the variational search in QAOA can be done with negligible computational cost.

There are other quantum algorithms that can also count ground states of some Hamiltonians, such as the quantum amplitude estimation algorithm and its variants~\cite{brassard1998quantum,brassard2002quantum,wie2019simpler}, and counting by sampling from the final wave function in a quantum algorithm~\cite{hen2014fast,aaronson2019quantum}. All of these rely on being able to implement Grover's oracle on a quantum circuit. Then, these algorithms can have a speedup over classical algorithms only if the classical Hamiltonians considered encode problems for which it is possible to verify if a given state is a solution to the problem in polynomial time, i.e., problems lying in the computational complexity class NP. Our algorithm is more general---it can be used to count weighted ground states of arbitrary Hamiltonians. Moreover, one of the techniques that we present, QAOA, has recently shown significant promise for implementation on NISQ devices and rapidly finding ground states. While we only observe a sub-quadratic speedup in our QAOA algorithm, further research might improve this speedup.

This article is organized as follows. In Sec.~\ref{sec: setup}, we define the ground-state counting problem we consider, and give a concrete example. In Secs.~\ref{subsec: Grover}--\ref{subsec: QAOA}, we describe modified quantum algorithms---Grover, AQO, and QAOA---for importance-sampling the ground states of the classical Hamiltonian. We calculate the scaling of the computational time for these algorithms, analytically in the case of AQO and Grover, and numerically in the case of QAOA. In Sec.~\ref{subsec: counting}, we describe a procedure to estimate the weighted count of ground states by iterating the experiment several times. In Sec.~\ref{sec: results}, we numerically compare the scaling of the total computational time required by our QAOA algorithm against classical OMCS, and show cases where QAOA scales more favorably with system size than OMCS. We summarize and provide a future outlook in Sec.~\ref{sec: summary}.

%% file: generalproblem.tex
The problem we consider in this work is estimating the total weighted count of ground states $\ket{g}$ of a classical Hamiltonian $\hH_z$ acting on a Hilbert space $\mathcal{H}$, with a nonnegative normalized weight function $w:\mathcal{H} \rightarrow [0,1]$, where $\sum_{\phi\in\mathcal{H}} w(\phi)=1$ with the sum running over the classical basis states of $\mathcal{H}$. The Hamiltonian can be general, with interactions between arbitrary numbers of spins,
\begin{equation}
\hH_z = \sum_{s\in P(1, \cdots, n)} J_s \prod_{j\in s} \hs^z_j,
\end{equation}
where $P(1,\cdots,n)$ is the powerset of $\{1,\cdots,n\}$, and $J_s$ are arbitrary real numbers.

We denote the distinct eigenvalues of $\hH_z$ as $\mathcal{E}_j$, where $0\leq j\leq m-1$, and let $\mathcal{E}_0 < \mathcal{E}_1 < \cdots \mathcal{E}_{m-1}$. Each eigenvalue can have degenerate eigenstates. We define moments, $N_j^{(\mu)}$, for the different manifolds as
\begin{equation}\label{eqn: moments}
N_j^{(\mu)} = \sum_{\phi: H_z(\phi)=\mathcal{E}_j} (w(\phi))^\mu.
\end{equation}
For notational convenience, we denote ground state moments, $N_0^{(\mu)}$, as $\mathcal{P}_\mu$.

The quantity we want to estimate---the total weighted count of the classical ground states $\ket{g}$ in the ground state space $\mathcal{G}$ of $\hH_z$---is
\begin{equation}
\mathcal{P} \equiv \mathcal{P}_1 = \sum_{g\in\mathcal{G}} w(g).
\end{equation}

%% file: edgecovers.tex
\subsection{An application: Edge covers}\label{subsec: edge covers}
As a concrete example of $\hH_z$, we consider counting edge covers on a graph, which is a local constraint-satisfaction problem. For a graph with vertices $v\in V$ and links $e\in E$, a subset $E' \subseteq E$ is said to be an edge cover if $E'$ has at least one link incident on every vertex in $V$. Figure~\ref{fig: edge covers}(a) illustrates some examples of edge covers and non-edge covers on the ``paw'' graph, also known as the 3-pan graph or the (3,1)-tadpole. 
The weighted count of edge covers is an upper-bound for the graph's all-terminal reliability~\cite{duenas2018quantum,paredes2018network}, which determines the probability that a graph stays connected when its links fail with a given probability. Efficiently calculating the all-terminal reliability has applications in designing reliable engineering systems~\cite{zio2009reliability}.

To recast counting weighted edge covers as a ground-state counting problem, we map each link to a qubit, and define a one-to-one map between every subset $E'$ and a state in a Hilbert space with $|E|$ bits. Every link in $E'$ is mapped to $\ket{0}$, and every link not in $E'$ is $\ket{1}$. Then, the set of edge covers forms a one-to-one mapping with the ground state space $\mathcal{G}$ of
\begin{equation}\label{eqn: Hz}
\hH_z = \sum_{v\in V} \prod_{e\in E(v)} \frac{1-\hs^z_e}{2},
\end{equation}
where $E(v)$ is the set of links $e$ incident on $v$, and $\hs^z = \ket{0}\bra{0}-\ket{1}\bra{1}$. For each node $v$, the product $\prod_{e\in E(v)} \frac{1-\hs^z_e}{2}$ is zero if any of the links incident on $v$ is $\ket{0}$ (i.e., present in $E'$), and is one if all the links incident on $v$ are $\ket{1}$ (i.e., none are present in $E'$). Therefore, the total energy of a classical state $\ket{E'}$ corresponding to a subset $E'$ is equal to the number of nodes $v$ not incident to any links in $E'$. The energy of all edge covers is $0$, and they form a one-to-one map with the ground states of $\hH_z$. The eigenvalues of $\hH_z$ for this problem are integers, $\mathcal{E}_j=j$ for $0\leq j\leq |V|-2$, and $\mathcal{E}_{|V|-1} = |V|$.

For this problem, we consider the weight on any state $\ket{E'}=\ket{e_1\cdots e_{|E|}}$ to be
\begin{equation}
w(E') = q^{n_1} (1-q)^{n_0}
\label{eqn: weights}\end{equation}
where $e_i\in\{0,1\}$, $q\in [0,1]$, and $n_1$ and $n_0$ are the number of $1$s and $0$s in $E'$. This weight naturally occurs in engineering applications where links fail independently with probability $q$.

Classical Monte Carlo algorithms give an estimate $\mathcal{P}_{\rm est}$ for the desired result $\mathcal{P}$ by importance-sampling the space of all link configurations (i.e., the powerset of $E$) with the probability distribution $w(E')$. Improved algorithms such as OMCS also provide a confidence $1-\delta$ on the relative error $\epsilon$, defined as
\begin{equation}\label{eqn: confidence}
1-\delta \equiv {\rm Pr}\left(\left|1-\frac{\mathcal{P}_{\rm est}}{\mathcal{P}}\right| < \epsilon\right).
\end{equation}
When $\epsilon, \delta \ll 1$, the number of samples drawn in OMCS to estimate $\mathcal{P}$ scales as~\cite{karp1983monte}
\begin{equation}\label{eqn: TOMCS}
T_{\rm OMCS}\sim |\ln(\delta)|/(\mathcal{P}\epsilon^2).
\end{equation}

%% file: methods_v2.tex
Our quantum algorithm to estimate $\mathcal{P}$ has two parts. In the first part, we coherently evolve the quantum system to a target wave function in the ground state space $\mathcal{G}$ and measure the system in the computational basis (Sec.~\ref{subsec: Grover}--\ref{subsec: QAOA}). In the second part of the algorithm, we iterate the first part several times, and do a classical statistical analysis on the measurements to estimate $\mathcal{P}$ (Sec.~\ref{subsec: counting}). The natural choice for a target wave function to count ground states $\ket{g}$ with weights $w(g)$ samples a ground state $\ket{g}$ with relative probability $w(g)$. This criterion is met by the choice
\begin{equation}\label{eqn: target}
\ket{\psi_{\rm target}} = \frac{1}{\sqrt{\mathcal{P}}} \sum_{g\in\mathcal{G}} \sqrt{w(g)}\ket{g},
\end{equation}
where $1/\sqrt{\mathcal{P}}$ is a normalization factor.

Although AQO and QAOA can often find the ground state space of a Hamiltonian faster than classical algorithms, they are unsuitable for reaching a pre-determined target wave function such as Eq.~(\ref{eqn: target}) in a degenerate space, when used with the usual mixing Hamiltonian $\hH_x = -\sum_{i=1}^n\hs_i^x$, and are therefore inefficient for counting ground states. For example, in the adiabatic limit of AQO, the final wave function is given by degenerate perturbation theory with $\hH_x$ as the perturbing term, and this wave function is not known \textit{a priori}. In fact, several works~\cite{boixo2013experimental,matsuda2009ground,konz2018uncertain,matsuda2009quantum,king2016degeneracy,mandra2017exponentially,zhang2017advantages,katzgraber2018viewing} have numerically found that  some ground states $\ket{g}$ are exponentially suppressed in the final wave function relative to other ground states $\ket{g'}$, so that $\left|\overlap{\psi(T)}{g}\right| \ll \left|\overlap{\psi(T)}{g'}\right|$. Finding the exponentially suppressed ground states by measuring the final wave function will require exponentially many experiments, and therefore, it becomes inefficient to count all the ground states. In QAOA, the distribution of classical ground states in the final wave function depends on the variational parameters used to evolve the system, and it is difficult to obtain confidence on estimates of the weighted count.

In this section, we solve the difficulties described above in using AQO and QAOA to count ground states of Hamiltonians. Specifically, we (i) modify AQO and QAOA to guarantee that the instantaneous wave function's amplitudes in the computational basis importance-sample the ground states, i.e., $\left|\overlap{\psi(t)}{g}\right|^2 \propto w(g)$, (ii) describe a statistical technique to count the ground states with weights $w(g)$, with a user-specified relative error and confidence, in the asymptotic limit of large system size, and (iii) analyze the asymptotic scaling of the computational time with problem size. Remarkably, besides enabling efficient counting, our modifications also allow us to analytically predict the asymptotic scaling of AQO.

Our modifications build on ideas proposed in Refs.~\cite{matsuda2009ground,hen2014fast,van2001powerful,roland2002quantum}, but our results are more general. Most importantly, while Refs.~\cite{hen2014fast,van2001powerful,roland2002quantum} analyzed AQO only for a restricted Hamiltonian $\hH_z = 1 - \sum_{g\in\mathcal{G}} \ket{g}\bra{g}$ (i.e., $e^{i\pi\hH_z}$ is a Grover oracle), our analytical results for AQO hold for arbitrary classical $\hH_z$, even those not easily implementable as Grover oracles. In this way, our work also opens avenues to solve counting problems that cannot be approached by the usual counting algorithms such as amplitude estimation~\cite{brassard1998quantum,brassard2002quantum,wie2019simpler}. Furthermore, while Ref.~\cite{matsuda2009ground} did not explicitly prove that their ideas lead to exactly fair sampling, we prove it, and we extend those ideas to importance-sampling. Our modifications still have close connections to Grover's algorithm, despite solving a larger class of problems, and therefore we will also briefly present the version of Grover's algorithm for weighted counting in Sec.~\ref{subsec: Grover}.

Our algorithms involve a few time scales. We denote the number of calls to $\hH_x$ and $\hH_z$ required to coherently evolve the system to $\mathcal{G}$ in one iteration of AQO and QAOA as $T_{\rm AQO}$ and $T_{\rm QAOA}$, and the number of oracle calls in Grover's algorithm as $T_{\rm Grover}$. Additionally, there is some overhead, $T_{\rm\alpha\beta\ search}$, for finding the variational parameters in QAOA. We give a rigorous statistical approach in Sec.~\ref{subsec: counting} to estimate $\mathcal{P}_{\rm est}$ with user-specified confidence on its relative error from the actual value $\mathcal{P}$, in the asymptotic limit of system size. We denote the number of iterations required for this statistical analysis as $T_{\rm count}$. The total times for the three algorithms then scale as $T_{\rm AQO} \times T_{\rm count}$, $T_{\rm\alpha\beta\ search} + T_{\rm QAOA} \times T_{\rm count}$, and $T_{\rm Grover} \times T_{\rm count}$. There are some overheads to this total time. For example, one source of a multiplicative overhead is the circuit to implement one discrete step of the quantum evolution in the first part of the algorithm. For the edge cover problem, this multiplicative overhead increases polynomially with the number of qubits. One of the additive overheads arises from determining $T_{\rm AQO}, T_{\rm QAOA}$, or $T_{\rm Grover}$ for evolving the system to $\mathcal{G}$. This overhead increases logarithmically with $T_{\rm AQO}, T_{\rm QAOA}$, and $T_{\rm Grover}$. For most practical problems and Hamiltonians of interest, both the multiplicative and additive overheads are subleading compared to $T_{\rm AQO}, T_{\rm QAOA}$, and $T_{\rm count}$, when $\mathcal{P}$ is exponentially small in the number of qubits. The overheads are subleading to $T_{\rm Grover}$ in Grover's algorithm for problems in NP.

%% file: grover.tex
Grover showed~\cite{grover1998quantum} that the target wave function $\ket{\psi_{\rm target}}$ in Eq.~(\ref{eqn: target}) can be reached in Grover's algorithm by choosing the initial state and the diffusion operator as
\begin{eqnarray}\label{eqn: Grover}
&\ket{\psi(0)} = \sum_\phi \sqrt{w(\phi)}\ket{\phi},\nonumber\\
&U_0 = 1 - 2\ket{\psi(0)}\bra{\psi(0)}.
\end{eqnarray}
The oracle is the same as usual, $U_\mathcal{G} = 2\hat{P}_\mathcal{G}-1$, where $\hat{P}_\mathcal{G} = \sum_{g\in\mathcal{G}} \ket{g}\bra{g}$ is the projector onto $\mathcal{G}$. Repeated iterations of $U_0U_\mathcal{G}$ rotate the wave function in the plane of $\ket{\psi(0)}$ and $\ket{\psi_{\rm target}}$, and the wave function reaches $\ket{\psi_{\rm target}}$ after $T_{\rm Grover}=\pi/(4\sqrt{\mathcal{P}})$ iterations.

Moreover, since the instantaneous wave function $\ket{\psi(t)}$ after $t$ Grover iterations is always a superposition of only $\ket{\psi(0)}$ and $\ket{\psi_{\rm target}}$, both of whose amplitudes in the computational basis importance-sample the ground states $\ket{g}$, the amplitudes of $\ket{\psi(t)}$ also importance-sample the ground states, up to an overall constant factor. That is,
\begin{equation} \label{eqn: importance-sampling}
\left|\frac{\overlap{\psi(t)}{g}}{\overlap{\psi(t)}{g'}}\right|^2 = \frac{w(g)}{w(g')} \ \forall\ t,\ \forall g,g'\in\mathcal{G}.
\end{equation}

The oracle $U_\mathcal{G}$ can be implemented with polynomially many gates (i.e., $n^r$ gates for $n$ qubits) on a  quantum circuit for Hamiltonians that encode classical problems in the computational complexity class NP. 
Polynomial-time implementations of Grover oracles do not exist for $\hH_z$ which encode problems outside NP. The complexity of the circuit for preparing the initial state $\ket{\psi(0)}$ and implementing $U_0$ depend on the function $w$. For the weights in Eq.~(\ref{eqn: weights}), 
$\ket{\psi(0)} = \bigotimes_{i=1}^{|E|} (\sqrt{1-q}\ket{0}_i + \sqrt{q}\ket{1}_i)$
is a product state.

Figures~\ref{fig: qtm cct}(a)-(c) show the circuit to prepare $\ket{\psi(0)}$ and implement $U_0$ and $U_\mathcal{G}$, for the problem defined in Eq.~(\ref{eqn: Hz}) and the weights in Eq.~(\ref{eqn: weights}). The state $\ket{\psi(0)}$ can be prepared with only single-qubit gates. Implementing $U_0$ and $U_\mathcal{G}$ require multi-qubit controlled-phase gates. There are several techniques to decompose the multi-qubit phase gates with $k$ bits to only one- and two-qubit gates, for example with $O(k)$ gates using $k-3$ ancillary bits~\cite{barenco1995elementary}, or $O(k^2)$ gates with no ancillary bits~\cite{barenco1995elementary,saeedi2013linear}.

%% file: aqo_v2.tex
AQO works by preparing the system in an initial state, which is also a ground state of a Hamiltonian $\hH_x$, and then adiabatically varying the Hamiltonian as $\hH(t) = \alpha(t)\hH_x + \beta(t)\hH_z$ from $t=0$ to $t=T_{\rm AQO}$, with $\alpha(0)=\beta(T_{\rm AQO})=1$ and $\alpha(T_{\rm AQO})=\beta(0)=0$. The most common choices for the initial state and the Hamiltonian are $\ket{\psi(0)} = \bigotimes_{i=1}^{|E|}\ (\ket{0}_i+\ket{1}_i)/\sqrt{2}$ and $\hH_x = -\sum_{i=1}^n\hs_i^x$. In some variations, $\beta$ is fixed while only the ratio $\alpha/\beta$ is varied from $\infty$ to $0$, which leads to the same final state as varying both $\alpha$ and $\beta$ with time. However, as has been observed before~\cite{boixo2013experimental,matsuda2009ground,konz2018uncertain,matsuda2009quantum,king2016degeneracy,mandra2017exponentially,zhang2017advantages,katzgraber2018viewing}, evolving with $\hH_x = -\sum_{i=1}^n\hs_i^x$ leads to exponential suppression of a significant number of classical ground states in the final wave function.

In this section, we will show that the final wave function $\ket{\psi_{\rm target}}$ can be reached in AQO by choosing the initial state as $\ket{\psi(0)}$ in Eq.~(\ref{eqn: Grover}), and the mixing Hamiltonian as
\begin{equation}\label{eqn: Hx}
\hH_x = \frac{U_0-1}{2} = - \ket{\psi(0)}\bra{\psi(0)},
\end{equation}
with $U_0$ in Eq.~(\ref{eqn: Grover}). We will also show that the amplitudes of the wave function in the computational basis, during any time of executing AQO, importance-sample the ground states of $\hH_z$. Both of these facts arise from the relation of $\hH_x$ to $U_0$. Therefore, like Grover's algorithm, the evolution of the wave function is restricted to lie in a smaller, symmetric, subspace than the full Hilbert space, and wave functions in this symmetric space importance-sample the ground states. 
The AQO schedule we consider is $\beta(t)=1-\alpha(t)=t/T_{\rm AQO}$. We analytically derive a lower bound for $T_{\rm AQO}$.

One can implement a discrete-time version of AQO on a circuit by applying the sequence of operators $\prod_{j=1}^{T_{\rm AQO}/\rmd t} \exp(-i\alpha(t_j)\hH_x\rmd t)\exp(-i\beta(t_j)\hH_z\rmd t)$ to $\ket{\psi(0)}$. Figures~\ref{fig: qtm cct}(c)-(d) show how to implement $\exp(i\alpha\hH_x)$ and $\exp(-i\beta\hH_z)$ for the paw graph in Fig.~\ref{fig: qtm cct}(a). One of the advantages of AQO (and QAOA in Sec.~\ref{subsec: QAOA}) is that it is possible to similarly construct circuits for $\exp(-i\beta \hH_z)$ with polynomially many (i.e., $n^r$ for $n$ qubits) gates for several other practical problems of interest outside NP, even when it is not possible to implement the Grover oracle with polynomially many gates.

If $\hH$ changes adiabatically, the adiabatic theorem guarantees that the final wave function at $t=T_{\rm AQO}$ will be a ground state of $\hH_z$. Specifically, the adiabatic theorem states that~\cite{messiah1964quantum}
\begin{equation}
\expectation{\psi(T_{\rm AQO})} {\hat{P}_\mathcal{G}} \geq 1-\eta^2
\end{equation}
if
\begin{equation} \label{eqn: adia thm}
 \left|\left|\rmd \hat{H}/\rmd t\right|\right| \leq \eta\Delta(t)^2,
\end{equation}
where $\Delta(t)$ is the instantaneous energy difference between the lowest two eigenstates of $\hH(t)$, and $||\cdots||$ denotes operator norm.

Next, we find the spectrum of $\hH(\alpha,\beta) = \alpha\hH_x + \beta\hH_z$, and use this to analyze the scaling of $T_{\rm AQO}$ with the system size for the adiabaticity condition to be satisfied. As an example, Fig.~\ref{fig: edge covers}(b) shows the spectrum of $\hH(\alpha/\beta,1)$ for the edge cover problem on the paw graph.

\subsubsection{Spectrum of $\hH(\alpha,\beta)$.}

The eigenstates of $\hH(\alpha,\beta)$ fall in two kinds. In the first kind, the eigenstates are anti-symmetric combinations $(\sqrt{w(\phi')}\ket{\phi} - \sqrt{w(\phi)}\ket{\phi'})/\sqrt{w(\phi)+w(\phi')}$, with eigenvalue $\lambda=\beta \mathcal{E}_j$, where both $\ket{\phi}$ and $\ket{\phi'}$ are classical states with  classical energy $H_z(\phi)=H_z(\phi')=\mathcal{E}_j$. For every $j$, there are $N_j^{(0)}-1$ such independent eigenstates of $\hH$. The eigenvalues of these states are shown as red lines in Fig.~\ref{fig: edge covers}(b). We will see that the wave function has no overlap with these eigenstates at any time during AQO or QAOA.

The second kind of eigenstates lie in a Hilbert space $\mathcal{H}_S$ spanned by the symmetric basis states
\begin{equation}\label{eqn: basis states}
\ket{\Phi_j} = \frac{\sum_{\phi: H_z(\phi) = \mathcal{E}_j} \sqrt{w(\phi)} \ket{\phi} }{\sqrt{N_j^{(1)}}}.
\end{equation}
Letting $\hat{P}_S$ be the projection operator into $\mathcal{H}_S$, the projected Hamiltonian is $\hH_S(\alpha,\beta) = \alpha\hH_{xS} + \beta\hH_{zS}$, where
\begin{eqnarray} \label{eqn: projected Hamiltonians}
&\hH_{zS} = \hat{P}_S\hH_z\hat{P}_S = \sum_j \mathcal{E}_j\ket{\Phi_j}\bra{\Phi_j}, \nonumber\\
&\hH_{xS} = \hat{P}_S\hH_x\hat{P}_S = - \left(\sum_i \sqrt{N_i^{(1)}}\ket{\Phi_i}\right) \left(\sum_j \sqrt{N_j^{(1)}}\bra{\Phi_j}\right).
\end{eqnarray}

The eigenvalue equation for $\hH_S$ is $\det(\hH_S(\alpha,\beta)-\lambda) = 0$. Note that $-\hH_{xS}$ is also a projection operator, like $-\hH_x$. Therefore, $\det(\hH_S(\alpha,\beta)-\lambda)$ is at most linear in $\alpha$, and we can use Taylor expansion and Jacobi's formula to write
\begin{eqnarray}\label{eqn: det}
\det(\hH_S(\alpha,\beta)-\lambda) &= \det(\beta\hH_{zS}-\lambda) + \alpha \frac{\rmd \det(\alpha\hH_{xS}+\beta\hH_{zS}-\lambda)}{\rmd\alpha} \nonumber\\
&= \det(\beta\hH_{zS}-\lambda) + \alpha \Tr(\hH_{xS}\ {\rm adj}(\beta\hH_{zS}-\lambda)),
\end{eqnarray}
where adj$(\cdots)$ is the adjugate. Substituting Eq.~(\ref{eqn: projected Hamiltonians}) into Eq.~(\ref{eqn: det}), we obtain
\begin{equation}\label{eqn: det2}
\det(\hH_S(\alpha,\beta)-\lambda) = \prod_j \left( \beta\mathcal{E}_j-\lambda \right) - \sum_k \alpha N_k^{(1)}\prod_{j\neq k} (\beta\mathcal{E}_k-\lambda).
\end{equation}
Then, the eigenvalues $\lambda$ of $\hH_S(\alpha,\beta)$ are given by the implicit algebraic equation
\begin{equation}\label{eqn: eigs}
\sum_j \frac{N_j^{(1)}} {\beta\mathcal{E}_j-\lambda} = \frac{1}{\alpha}.
\end{equation}
For $\alpha, \beta, N_{j-1}^{(1)}, N_j^{(1)} > 0$, the left hand side of this equation is a function of $\lambda$ which monotonically increases from $-\infty$ to $\infty$ as $\lambda$ changes from $\beta\mathcal{E}_{j-1}$ to $\beta\mathcal{E}_j$. Therefore, Eq.~(\ref{eqn: eigs}) has exactly one solution in the range
\begin{eqnarray}\label{eqn: Erange}
& \beta\mathcal{E}_{j-1} \leq \lambda_j \leq \beta\mathcal{E}_j,\ 0<j<m,\nonumber\\
& \beta\mathcal{E}_0-\alpha \leq \lambda_0 \leq \beta\mathcal{E}_0.
\end{eqnarray}
The equalities, $\lambda_j = \beta\mathcal{E}_j$ or $\lambda_{j+1}=\beta\mathcal{E}_j$, hold true only when $\alpha N_j^{(1)}=0$, and $\lambda_0=\beta\mathcal{E}_0-\alpha$ only when $N_0^{(1)}=\mathcal{P}=1$.

\subsubsection{Proof of importance-sampling.}

$\mathcal{H}_S$ is closed under the action of unitaries $\exp(-i\hH_x\alpha)$ and $\exp(-i\hH_z\beta)$, for arbitrary $\alpha$ and $\beta$. The initial state $\ket{\psi(0)}$ [in Eq.~(\ref{eqn: Grover})] lies in $\mathcal{H}_S$, and therefore the instantaneous wave function during any time in AQO lies in $\mathcal{H}_S$. As a result, the instantaneous wave function always importance-samples the ground states $\ket{g}$, leading to Eq.~(\ref{eqn: importance-sampling}) for AQO as well. This is illustrated in Fig.~\ref{fig: main results}(a).

In the adiabatic limit, $\ket{\psi(T_{\rm AQO})}$ lies in $\mathcal{G}$ and in $\mathcal{H}_S$, therefore $\ket{\psi(T_{\rm AQO})} = \ket{\psi_{\rm target}}$.

\subsubsection{Calculating $T_{\rm AQO}$.} 

Naively, one expects that the evolution time $T_{\rm AQO}$ for the adiabaticity condition [Eq.~(\ref{eqn: adia thm})] to be satisfied is $T_{\rm AQO}=\infty$. This naive expectation is because $\hH$ has ground state degeneracy at $t=T_{\rm AQO}$, and therefore the minimum energy gap $\Delta^*$ above $\ket{\psi_{\rm target}}$ is $0$. However, the required $T_{\rm AQO}$ for Eq.~(\ref{eqn: adia thm}) to be satisfied is in fact finite, because the instantaneous wave function always lies in $\mathcal{H}_S$. The energy gap in this subspace is $\Delta = \lambda_1-\lambda_0 > 0$ always if $\mathcal{P}>0$, as shown by Eqs.~(\ref{eqn: eigs}) and~(\ref{eqn: Erange}).

We can lower-bound $T_{\rm AQO}$ for the adiabatic condition to be met, by estimating the minimum value of $\Delta$.  We find from Eq.~(\ref{eqn: det2}) that
\begin{equation}\label{eqn: prod}
\prod_j (\lambda_j - \beta \mathcal{E}_0) = \det(\hH_S(\alpha,\beta) - \beta\mathcal{E}_0) = -\alpha \mathcal{P} \prod_{j\neq0} \beta(\mathcal{E}_j-\mathcal{E}_0).
\end{equation}
Eqs.~(\ref{eqn: Erange}) and~(\ref{eqn: prod}) then result in the inequality
\begin{equation}\label{eqn: inequality}
\alpha\beta\mathcal{P}(\mathcal{E}_1 -\mathcal{E}_0) \leq (\beta\mathcal{E}_0-\lambda_0) (\lambda_1-\beta\mathcal{E}_0) \leq \alpha\beta\mathcal{P} (\mathcal{E}_{m-1}-\mathcal{E}_0).
\end{equation}
Using the relation $|x+y| \geq 2\sqrt{xy}$, and setting $x=\lambda_1-\beta\mathcal{E}_0, y=\beta\mathcal{E}_0-\lambda_0$, we obtain
\begin{equation}
\Delta = \lambda_1-\lambda_0 \geq 2\sqrt{\alpha\beta\mathcal{P}(\mathcal{E}_1-\mathcal{E}_0)}.
\end{equation}

Next, we obtain bounds for $\beta=\beta^*$ and $\alpha=1-\beta^*$ where the minimum value $\Delta=\Delta^*$ occurs, for $\mathcal{P} \ll 1$. We will assume that $\mathcal{E}_1-\mathcal{E}_0 \gtrsim O(1)$ and $\Delta^* \ll 1$.  The latter is typically valid when $\mathcal{P}\ll1$ and $\mathcal{E}_1-\mathcal{E}_0 \gtrsim O(1)$. The sum of eigenvalues of $\hH_s(\alpha,\beta)$ is $\sum_j \lambda_j = \Tr(\hH_s(\alpha,\beta)) = -\alpha + \sum_j \beta\mathcal{E}_j$. When this is combined with the inequalities for $\lambda_2,\cdots,\lambda_{m-1}$ in Eq.~(\ref{eqn: Erange}), we find that $\beta(\mathcal{E}_1+\mathcal{E}_0) - \alpha \leq \lambda_1 + \lambda_0 \leq \beta(\mathcal{E}_{m-1}+\mathcal{E}_0) - \alpha$, which can be rewritten as
\begin{equation}\label{eqn: lambda1+lambda0 inequality}
\beta(\mathcal{E}_1-\mathcal{E}_0)-\alpha \leq (\lambda_1 - \beta\mathcal{E}_0) - (\beta\mathcal{E}_0 - \lambda_0) \leq \beta(\mathcal{E}_{m-1}-\mathcal{E}_0)-\alpha.
\end{equation}
For $\beta(\mathcal{E}_1-\mathcal{E}_0) \gg \alpha$, the first inequality in Eq.~(\ref{eqn: lambda1+lambda0 inequality}) can be satisfied only if $\beta\mathcal{E}_0 - \lambda_0 \ll \lambda_1-\beta\mathcal{E}_0 \sim \beta(\mathcal{E}_1-\mathcal{E}_0)$. In this limit, $\Delta \sim \beta(\mathcal{E}_1-\mathcal{E}_0) \gtrsim O(1)$, which is much larger than the minimum value it can take, $\sqrt{4\alpha\beta\mathcal{P}(\mathcal{E}_1-\mathcal{E}_0)}$, since $\beta(\mathcal{E}_1-\mathcal{E}_0)/\sqrt{4\alpha\beta\mathcal{P}(\mathcal{E}_1-\mathcal{E}_0)} \gg 1/\sqrt{4\mathcal{P}} \gg 1$. 
For $\beta(\mathcal{E}_{m-1}-\mathcal{E}_0) \ll \alpha$, the second inequality in Eq.~(\ref{eqn: lambda1+lambda0 inequality}) can be satisfied only if $\lambda_1-\beta\mathcal{E}_0 \ll \beta\mathcal{E}_0-\lambda_0 \sim \alpha$. In this limit, $\Delta \sim \alpha \sim 1$, which is again much larger than the minimum value it can take, since $\alpha/\sqrt{4\alpha\beta\mathcal{P}(\mathcal{E}_1-\mathcal{E}_0)} > \sqrt{(\mathcal{E}_{m-1}-\mathcal{E}_0)/(4\mathcal{P}(\mathcal{E}_1-\mathcal{E}_0))} \gg 1$. 
Then, $\alpha^*$ and $\beta^*$ do not lie in either of the two limits above, leading to
\begin{eqnarray}\label{eqn: alpha* and beta*}
& \beta^*(\mathcal{E}_1 - \mathcal{E}_0) \lesssim \alpha^* \lesssim \beta^*(\mathcal{E}_{m-1} - \mathcal{E}_0), \nonumber\\
\Rightarrow & \frac{1}{1+\mathcal{E}_{m-1}-\mathcal{E}_0} \lesssim \beta^* \lesssim \frac{1}{1+\mathcal{E}_1-\mathcal{E}_0}.
\end{eqnarray}
The minimum value of $\Delta^*$ depends on the product $\beta^*(1-\beta^*)$. Since the function $f(\beta)=\beta(1-\beta)$ has no local minima, $\min_{x\leq \beta\leq y}f(\beta) = \min(x(1-x),y(1-y))$. That is, for $\beta^*$ lying in the interval given by Eq.~(\ref{eqn: alpha* and beta*}),
\begin{equation}\label{eqn: Delta}
\Delta^* \geq \left(4(\mathcal{E}_1-\mathcal{E}_0)\min\left( \frac{\mathcal{E}_1-\mathcal{E}_0}{(1+\mathcal{E}_1-\mathcal{E}_0)^2}, \frac{\mathcal{E}_{m-1}-\mathcal{E}_0}{(1+\mathcal{E}_{m-1}-\mathcal{E}_0)^2} \right)\right)^{1/2} \sqrt{\mathcal{P}}.
\end{equation}
Then, using Eq.~(\ref{eqn: adia thm}) and the relation $||d\hH/\rmd t|| = ||\hH_z - \hH_x||/T_{\rm AQO} \leq (\mathcal{E}_{m-1}+1)/T_{\rm AQO}$,
\begin{equation}\label{eqn: TAQO full}
T_{\rm AQO} \geq \frac{1}{4\eta\mathcal{P}} \frac{\mathcal{E}_{m-1}+1}{(\mathcal{E}_1-\mathcal{E}_0) \min\left( \frac{\mathcal{E}_1-\mathcal{E}_0}{(1+\mathcal{E}_1-\mathcal{E}_0)^2}, \frac{\mathcal{E}_{m-1}-\mathcal{E}_0}{(1+\mathcal{E}_{m-1}-\mathcal{E}_0)^2} \right) }.
\end{equation}
This is a generalization of the result found in Refs.~\cite{hen2014fast,roland2002quantum,van2001powerful}, extended to importance-sample ground states of a general classical Hamiltonian with a general weight function. Our result has additional factors arising from $\alpha^*\beta^*$ and $||\hH_z - \hH_x||$ (which were $1/4$ and $2$ respectively in~\cite{hen2014fast,roland2002quantum,van2001powerful}). When our assumption is violated, i.e. $\mathcal{E}_1-\mathcal{E}_0 \ll 1$, bounds similar to Eq.~(\ref{eqn: alpha* and beta*}),~(\ref{eqn: Delta}) and~(\ref{eqn: TAQO full}) can be derived by changing the AQO schedule to $\alpha = (\mathcal{E}_1 - \mathcal{E}_0)(1-\beta)$.

For the edge cover problem, $\mathcal{E}_0=0, \mathcal{E}_1=1$ and $\mathcal{E}_{m-1}=|V|$. Then, Eq.~(\ref{eqn: TAQO full}) gives
\begin{equation}\label{eqn: TAQO}
T_{\rm AQO} \geq \frac{(|V|+1)^3}{4\eta|V|\mathcal{P}}.
\end{equation}

\subsubsection{Numerical simulation of AQO for edge covers.}

Figures~\ref{fig: AQO and qaoa times}(a)-(b) numerically confirm the scaling in Eq.~(\ref{eqn: TAQO}), and the applicability of this asymptotic formula for finite problem sizes. They plot the number of discrete AQO steps $T_{\rm AQO}/\rmd t$ required to reach $\expectation{\psi(T_{\rm AQO})}{P_\mathcal{G}} = 1-\eta^2$ in a simulation of discrete-time AQO of $\hH_z$ in Eq.~(\ref{eqn: Hz}), with discrete time intervals $\rmd t$. In Fig.~\ref{fig: AQO and qaoa times}(a), we plot $T_{\rm AQO}/\rmd t$ for linear graphs with $q=\sin^2(0.3\pi)$, and in Fig.~\ref{fig: AQO and qaoa times}(b), for an ensemble of random graphs of different vertex degrees and different weighting parameters $q$. In both these cases, we find that $T_{\rm AQO}\sim 1/\mathcal{P}$. We did not verify the logarithmic corrections to this scaling, $(|V|+1)^3/|V|$, in Eq.~(\ref{eqn: TAQO}). We arbitrarily chose $\eta$ and $\rmd t$ for these plots, but we find the same scaling for any $\eta$ and small enough $\rmd t$.

$T_{\rm AQO}$ in Eq.~(\ref{eqn: TAQO}) scales the same way as $T_{\rm OMCS}$ for fixed $\epsilon$ and $\delta$ [see Eq.~(\ref{eqn: TOMCS})]. Because of the additional counting overhead $T_{\rm count}$ that will be described in Sec.~\ref{subsec: counting}, the total time taken by AQO, $T_{\rm AQO}\times T_{\rm count}$, increases faster with system size than OMCS. Alternative AQO schedules, such as the one in Refs.~\cite{hen2014fast,roland2002quantum,van2001powerful} where the functional forms of $\alpha(t)$ and $\beta(t)$ are optimally chosen, could result in a quadratic speedup of $T_{\rm AQO}$. Rather than pursuing this, we next use a variational algorithm, QAOA, to optimize the quantum evolution. We find potential for a quadratic speedup.

%% file: qaoa.tex
QAOA is a classical-quantum hybrid variational algorithm that achieves the same goal as AQO, but has a circuit depth that scales more favorably with system size, if the angles $\alpha_j$ and $\beta_j$ (defined below) are chosen optimally as $\alpha_j^{\rm opt}$ and $\beta_j^{\rm opt}$ at each time step $t_j$. In this hybrid algorithm, one performs a quantum evolution with a certain choice for $\alpha_j$ and $\beta_j$, and evaluates a metric such as $\expectation{\psi(j)}{ \hat{P}_\mathcal{G} }$ by measuring the qubits in the computational basis at the end of the evolution. One then uses calls to this quantum algorithm from classical routines to find the best values $\alpha_j^{\rm opt}$ and $\beta_j^{\rm opt}$ that maximize this metric with the smallest number of time steps, $T_{\rm QAOA}$, required to reach sufficiently large $\expectation{\psi(T_{\rm QAOA})}{ \hat{P}_\mathcal{G} }$. If this metric cannot be implemented easily, e.g. for $\hH_z$ that encodes problems outside NP so its ground states are not verifiable in polynomial time, one could use a different metric that is easier to implement. An example of such a metric is $\expectation{\psi(j)}{  \hH_z }$, in cases where it is easier to implement than $\expectation{\psi(j)}{ \hat{P}_\mathcal{G} }$. Note that $\expectation{\psi(j)}{\hat{P}_\mathcal{G}} = |\overlap{\psi(j)}{\psi_{\rm target}}|^2$, since $\ket{\psi(j)}$ still lies in the symmetric subspace $\mathcal{H}_S$ if $\ket{\psi(0)}$ and $\hH_x$ are chosen as given in Eqs.~(\ref{eqn: Grover}) and~(\ref{eqn: Hx}).

Here, we consider two variational search methods to find $\alpha_j^{\rm opt}$ and $\beta_j^{\rm opt}$ for the edge cover problem. 
First, we do a greedy method where, for $\ket{\psi(j)}$ recursively defined as $\ket{\psi(j)} = \exp(-i\alpha_j\hH_x)\exp(-i\beta_j\hH_z)\ket{\psi(j-1)}$, $\alpha_j = \alpha_j^{\rm opt}$ and $\beta_j = \beta_j^{\rm opt}$ are chosen to maximize $\expectation{\psi(j)}{ \hat{P}_\mathcal{G} }$ for fixed $\alpha_1,\cdots,\alpha_{j-1},\beta_1,\cdots,\beta_{j-1}$. The points in Fig.~\ref{fig: qaoa with constant angles} show the results of numerically implementing the greedy method for a random instance of the edge cover problem shown in the inset of Fig.~\ref{fig: qaoa with constant angles}(b) with $q = \sin^2(0.3\pi)$. We evolve the system until it reaches $\expectation{\psi(j)}{\hat{P}_\mathcal{G}} = 0.5$. Figure~\ref{fig: qaoa with constant angles}(a) plots $\alpha_j^{\rm opt}$ and $\beta_j^{\rm opt}$ versus $j$, and Fig.~\ref{fig: qaoa with constant angles}(b) plots $\expectation{\psi(j)}{ \hat{P}_\mathcal{G} }$.

To analyze the scaling of the circuit depth in the greedy method versus problem size, we repeat this procedure for a larger variety of graphs and weights. Figure~\ref{fig: AQO and qaoa times}(a) shows $T_{\rm QAOA}$ required to reach $\expectation{\psi(T_{\rm QAOA})}{ \hat{P}_\mathcal{G} }= 0.8 $ for linear graphs at $q=\sin^2(0.3\pi)$, and Fig.~\ref{fig: AQO and qaoa times}(c) shows $T_{\rm QAOA}$ required to reach $\expectation{\psi(T_{\rm QAOA})}{ \hat{P}_\mathcal{G} }= 0.5 $ for the same random ensemble of graphs and weighting parameters $q$ used in Fig.~\ref{fig: AQO and qaoa times}(b). We observe that
\begin{equation}
T_{\rm QAOA} \sim 1/\sqrt{\mathcal{P}},
\end{equation}
possibly up to logarithmic corrections. This is the same scaling as the number of Grover iterations in Grover's algorithm, $T_{\rm Grover} \sim 1/\sqrt{\mathcal{P}}$. There is a greater spread of $T_{\rm QAOA}$ versus $1/\sqrt{\mathcal{P}}$ than $T_{\rm Grover}$ versus $1/\sqrt{\mathcal{P}}$ or $T_{\rm AQO}$ versus $1/\mathcal{P}$, however, we observe that $1 < T_{\rm QAOA}/T_{\rm Grover} < 2$ nearly always, even when $\mathcal{P}$ changes by $6$ orders of magnitude in Fig.~\ref{fig: AQO and qaoa times}(c).

In addition to the circuit depth, the greedy method involves another time scale--- the time to find the variational parameters $\alpha_j^{\rm opt}$ and $\beta_j^{\rm opt}$. Since finding $\alpha_j^{\rm opt}$ and $\beta_j^{\rm opt}$ at the $j$th step in the greedy method requires preparing $\ket{\psi(j-1)}$, the time required to find $\alpha_j^{\rm opt}$ and $\beta_j^{\rm opt}$ must scale as at least $O(j)$. Therefore, the total time $T_{\rm \alpha\beta\ search}^{\rm greedy}$ to find $\alpha_j^{\rm opt}$ and $\beta_j^{\rm opt}$ for $j=1,\cdots, T_{\rm QAOA}$ in the greedy method scales as 
$T_{\rm \alpha\beta\ search}^{\rm greedy} \sim T^2_{\rm QAOA}$.

The motivation for our second variational search method is to reduce $T_{\rm\alpha\beta\ search}$. Our second method is based on a simple observation about $\alpha_j^{\rm opt}$ and $\beta_j^{\rm opt}$ in the greedy method in Fig.~\ref{fig: qaoa with constant angles}(a)---they are nearly constant with $j$. Based on this, we propose fixing $\alpha_j^{\rm opt}$ and $\beta_j^{\rm opt}$ at constant values. We note that this trend of nearly constant $\alpha_j^{\rm opt}$ and $\beta_j^{\rm opt}$ occurs for most of the edge cover problems, but not necessarily all of them.

The solid lines in Fig.~\ref{fig: qaoa with constant angles} show the results of numerically implementing our second QAOA method, for the same random graph as the greedy method, but with constant $\alpha^{\rm opt}=0.78\pi$ and $\beta^{\rm opt}=0.12\pi$. Remarkably, $\expectation{\psi(j)}{ \hat{P}_\mathcal{G} }$ in Fig.~\ref{fig: qaoa with constant angles}(b) varies nearly identically when we use these constant parameters as it did with the greedily obtained parameters. Most importantly, the time required to variationally find constant $\alpha^{\rm opt}$ and $\beta^{\rm opt}$ is 
$T_{\rm\alpha\beta\ search}^{\rm constant} \sim O(1)$.

The scaling $T_{\rm QAOA} \sim T_{\rm Grover}$ is not surprising, because QAOA with our mixing Hamiltonian has close connections to Grover's algorithm. The unitary $\exp(-i\alpha\hH_x)$ is linearly related to the diffusion operator $U_0$, and is equal to $U_0$ for $\alpha=\pi$. The unitary $\exp(-i\beta\hH_z)$, which multiplies different energy manifolds of $\hH_z$ by different phases, is a generalization of the oracle $U_\mathcal{G}$ which multiplies all excited states by $-1$. In fact, we even find cases where QAOA is identical to Grover's algorithm. Two examples are the triangle graph and the linear $1\times3$ graph. For both these graphs, the ground state energy of $H_z$ is $0$, and the excited energies are odd numbers, $H_z=1$ or $3$. Therefore, the optimal QAOA parameters are $\alpha_j^{\rm opt} = \beta_j^{\rm opt} = \pi$, and $\exp(-i\alpha_j^{\rm opt}\hH_x) = U_0, \exp(-i\beta_j^{\rm opt}\hH_z) = U_\mathcal{G}$.

The total time taken by QAOA to estimate $\mathcal{P}$ scales as $T_{\rm \alpha\beta\ search} + T_{\rm QAOA} \times T_{\rm count}$. The greedy method, which has $T_{\rm \alpha\beta\ search}\sim T_{\rm QAOA}^2 \sim 1/\mathcal{P}$, has no speedup over OMCS, whose computational time also scales as $T_{\rm OMCS}\sim 1/\mathcal{P}$. However, in many cases, we find that $\alpha_j^{\rm opt}$ and $\beta_j^{\rm opt}$ are nearly constant and therefore it is possible to find the optimal parameters in $T_{\rm \alpha\beta\ search}\sim O(1)$. In this case, QAOA has a sub-quadratic speedup over OMCS, as we will see in Sec.~\ref{sec: results}. Motivated by this, in the rest of this paper, we show results for $T_{\rm QAOA}$ obtained from the greedy method, and neglect the overhead $T_{\alpha\beta\rm\ search}$ for finding $\alpha_j^{\rm opt}$ and $\beta_j^{\rm opt}$. Finding such quick variational optimization routines with small $T_{\alpha\beta\rm\ search}$ is an ongoing area of research~\cite{zhou2018quantum,morales2018variational,wecker2016training,mbeng2019quantum,parrish2019jacobi,niu2019optimizing,guerreschi2017practical,harrow2019low,gilyen2019optimizing}.

%% file: counting.tex
\subsection{Counting solutions by repeated measurements}\label{subsec: counting}
In this section, we show how to estimate $\mathcal{P}$, with a user-specified confidence on its relative error, in the asymptotic limit of large system size. We do this by iterating either of the three algorithms described in Secs.~\ref{subsec: Grover}-\ref{subsec: QAOA} $T_{\rm count}$ times, and analyzing the measurements in those experiments using the capture-recapture method~\cite{seber1973estimation,seber1986review}, generalized to count with weights $w(g)$. We describe this procedure below.

After evolving the system to a state $\ket{\psi(T)}$ with a large overlap, $1-\eta^2$, with $\mathcal{G}$, it is measured in the computational basis, giving us a ground state of $\hH_z$ with probability $\expectation{\psi(T)}{P_\mathcal{G}}=1-\eta^2$. Let $M$ denote the number of times a ground state is measured in $M'$ iterations. We statistically analyze only these $M$ states to estimate $\mathcal{P}$, and discard all the excited states measured. For problems outside NP, where one cannot verify when a ground state is measured in polynomial time, $M$ could denote the number of states measured with the lowest $H_z$ and therefore assumed to be ground states. This is a weaker criterion than counting states which are certain to be ground states.

Of the $M$ ground states measured, we denote the number of \textit{distinct} ground states measured as $Q_M$, and the total weight $w(g)$ of \textit{all} the ground states measured as $R_M$. Both $Q_M$ and $R_M$ are sharply peaked random variables with mean (see~\ref{sec: QMRM})
\begin{eqnarray}\label{eqn: means}
\mean{Q_M} =& \sum_{\mu=1}^M (-1)^{\mu-1} \left(\begin{array}{c}M\\ \mu\end{array}\right) \frac{\mathcal{P}_\mu} {\mathcal{P}^\mu}, \nonumber\\
\mean{R_M} =& M\frac{\mathcal{P}_2}{\mathcal{P}}.
\end{eqnarray}
In the limit that $\mathcal{P}_\mu / \mathcal{P}^\mu $ rapidly decays with $\mu$, which is true for large problem instances at fixed $q\notin\{0,1\}$, the series for $\mean{Q_M}$ can be truncated at $O(\mathcal{P}_2/\mathcal{P}^2)$, giving
\begin{equation}\label{eqn: QM approx mean}
\mean{Q_M} \approx M - \frac{M(M-1)}{2} \frac{\mathcal{P}_2} {\mathcal{P}^2}.
\end{equation}
$\mathcal{P}$ can then be obtained from Eqs.~(\ref{eqn: means}) and~(\ref{eqn: QM approx mean}) as
\begin{equation}\label{eqn: formula for P}
\mathcal{P} \approx \frac{ M(M-1)\mean{R_M} }{ 2(M-\mean{Q_M}) }.
\end{equation}

In practice, one would estimate $\mathcal{P}_{\rm est}$ by making $S$ measurements of $Q_M$ and $R_M$ and finding the sample means $\overline{R}_M$ and $\overline{Q}_M$ from this sample of size $M\times S$. This estimate for $\mathcal{P}$ would have some relative error to the actual $\mathcal{P}$. For sufficiently large $S$ used to estimate $\overline{Q}_M$ and $\overline{R}_M$, the relative error can be upper-bounded by $\epsilon$ with confidence $1-\delta$ [Eq.~(\ref{eqn: confidence})] by appealing to the central limit theorem. The confidence is given by  (see~\ref{sec: confidence})
\begin{equation}\label{eqn: delta}
1-\delta 
= \frac{1}{2}{\rm erf}\left( \frac{\epsilon}{1-\epsilon}\sqrt{\frac{M(M-1)S\mathcal{P}_2}{2\mathcal{P}^2}} \right)
 + \frac{1}{2}{\rm erf}\left( \frac{\epsilon}{1+\epsilon}\sqrt{\frac{M(M-1)S\mathcal{P}_2}{2\mathcal{P}^2}} \right).
\end{equation}
Inverting this relation for large $S$---so that the central limit theorem applies---and for $\epsilon,\delta \ll 1$, we obtain
\begin{equation}
S \gtrsim O \left( \frac{ |\ln(\delta)| \mathcal{P}^2 }{ \mathcal{P}_2 M^2\epsilon^2} \right).
\end{equation}

The total number of iterations in this procedure, $T_{\rm count} = M'\times S \sim M\times S/(1-\eta^2)$, is minimized by maximizing $M$. However, $M$ cannot be increased indefinitely, since $S$ has to be a large enough integer. We let $S\sim O(1)$, resulting in
\begin{equation}
T_{\rm count} \sim \frac{M}{1-\eta^2} \sim \frac{\sqrt{|\ln(\delta)|}}{\epsilon(1-\eta^2)} \sqrt{\frac{ \mathcal{P}^2 }{ \mathcal{P}_2 }}.
\end{equation}
It is noteworthy that $T_{\rm count} \propto \sqrt{|\ln(\delta)|}/\epsilon$ scales more favorably with $\epsilon$ and $\delta$, as compared to $T_{\rm OMCS}$ which scales as $|\ln(\delta)|/\epsilon^2$.

%% file: results_v2.tex
\section{Results: comparing gate counts in QAOA and OMCS}\label{sec: results}

The full quantum algorithm, including subroutines for determining the number of time steps in each iteration to evolve the system close to $\mathcal{G}$, recording ground states of $\hH_z$, and doing the statistical analysis on the measured ground states to estimate $\mathcal{P}$, is presented in~\ref{sec: full algo}. For brevity, we only describe AQO for the edge cover problem in~\ref{sec: full algo}. QAOA and Grover's algorithm can be implemented in a similar fashion.

The total number of one- and two-qubit gates in our quantum algorithms to estimate the weighted count $\mathcal{P}$, with confidence $1-\delta$ on the maximum relative error $\epsilon$, is asymptotically $T_{\rm count} \times (T_{\psi(0)} + (T_x + T_z)\times {\rm number\ of\ time\ steps})$, where the number of time steps is $T_{\rm Grover}, T_{\rm AQO}/\rmd t$ or $T_{\rm QAOA}$. The scaling of the number of steps in one iteration, and the number of iterations $T_{\rm count}$, is shown in Table~\ref{tab}. The scaling for $T_{\rm QAOA}$ is numerically observed for the edge cover problem, while $T_{\rm AQO}/\rmd t$ and $T_{\rm Grover}$ were analytically derived. $T_{\psi(0)}, T_x$, and $T_z$ are the number of gates required to prepare $\ket{\psi(0)}$, and to implement $\exp(i\alpha\hH_x)$, and $\exp(i\beta\hH_z)$, in AQO and QAOA. For Grover's algorithm, $T_x$ and $T_z$ refer to the number of gates required to implement $U_0$ and $U_\mathcal{G}$. In addition to the gate counts described here, there are additional overheads, such as $T_{\alpha\beta\rm\ search}$ for finding the variational parameters in QAOA, and trial experiments for finding $T_{\rm AQO}$ or $T_{\rm Grover}$ as described in~\ref{sec: full algo}. In heuristic methods like the one described in Sec.~\ref{subsec: QAOA}, where $\alpha^{\rm opt}$ and $\beta^{\rm opt}$ are constant, $T_{\alpha\beta\rm\ search}\sim O(1)$. Some of the other overheads were discussed in Sec.~\ref{sec: methods}.

There is a close correspondence between the number of quantum gates used by our algorithm to implement one discrete step of quantum evolution, and the computational times in OMCS for one random sample. The number of gates $T_z$ to implement $\exp(i\beta\hH_z)$ scales the same way as calculating $H_z$ for a random classical state, for arbitrary $\hH_z$, and both scale as $|E|$ for the edge cover problem if ancillary qubits are used. For $\hH_z$ that encodes problems in NP, $T_z$ for Grover's algorithm scales the same way as $T_z$ for AQO and QAOA. There can be a polynomial overhead to implement multi-qubit gates if no ancillary qubits are used. Similarly, $T_x$ and $T_{\psi(0)}$ scale the same way as the computational time for drawing one random sample in OMCS, for an arbitrary distribution $w(\phi)$, plus polynomial overheads for implementing multi-qubit gates.

Figures~\ref{fig: main results}(b-d) show the scaling of the total computational time taken by OMCS and QAOA to estimate $\mathcal{P}$ with $\epsilon=0.05$ and $\delta=0.05$, obtained from a numerical simulation of these two algorithms. We consider linear graphs with $q=\sin^2(0.35\pi)$ in Fig.~\ref{fig: main results}(b), and graphs of type $2\times n$ with $q=\sin^2(0.4\pi)$ and $q=\sin^2(0.35\pi)$ in Figs.~\ref{fig: main results}(c) and~(d). For OMCS, the computational time is the physical CPU time in seconds. For QAOA, the computational ``time'' refers to the total number of one- and two-qubit gates, i.e., $T_{\rm count}\times (T_{\psi(0)} + (T_x+T_z)\times T_{\rm QAOA})$, multiplied by a constant factor to lie on the same scale as OMCS. The gates in $T_x, T_z$ and $T_{\psi(0)}$ are counted assuming $|E|-3$ ancillary bits are used, so that multi-qubit gates are implemented using $O(|E|)$ one-qubit gates and CNOTs. Only the asymptotically leading terms are included, and we have neglected $T_{\rm\alpha\beta\ search}$ and the time required to find the depth $T_{\rm QAOA}$. The total ``time'' for QAOA in Fig.~\ref{fig: main results} also does not include classical overheads incurred for e.g. the statistical analysis.

For all the three cases shown in Figs.~\ref{fig: main results}(b)-(d), the total time for OMCS increases faster than it does for QAOA with problem size. The speedup in QAOA is sub-quadratic. We do not show the time for AQO, because it is not faster than OMCS. The computational time for Grover's algorithm scales the same way as QAOA.

It is worth noting that for all the graphs considered in Figs.~\ref{fig: main results}(b)-(d), $\mathcal{P}$ can be exactly computed in polynomial time $O(|E|^r)$ with classical algorithms. For the case of linear graphs, $\mathcal{P}\equiv\mathcal{P}_1$ is even a special case of a family of analytically known formulae:
\begin{equation}\label{eqn: analyticalP}
\mathcal{P}_\mu = \sum_r \left(\begin{array}{c}|E|-r-1 \\ r\end{array}\right) q^{\mu r} (1-q)^{\mu(|E|-r)},
\end{equation}
obtained from the recursive relation $\mathcal{P}_\mu(|E|) = q^\mu\mathcal{P}_\mu(|E|-2) + (1-q)^\mu\mathcal{P}_\mu(|E|-1)$. In particular, $\mathcal{P}_0(|E|) = {\rm Fib}(|E|)$ is a Fibonacci number. More general graphs do not have such closed-form formulae or polynomial-time algorithms, and OMCS or brute force are the best available classical choices.

The quantum advantage in QAOA, observed in Fig.~\ref{fig: main results} for grid graphs, arises from the quadratic speedup $T_{\rm QAOA} \sim 1/\sqrt{\mathcal{P}}$. The total computational time for QAOA, which scales as $T_{\alpha\beta\rm\ search}+T_{\rm QAOA}\times T_{\rm count}$, is asymptotically lesser than that for OMCS, despite including the multiplicative overhead $T_{\rm count} \propto \sqrt{\mathcal{P}^2/\mathcal{P}_2}$. As shown in Fig.~\ref{fig: AQO and qaoa times}, $T_{\rm QAOA}$ has a quadratic speedup even for random graphs and different weighting parameters. Therefore, we expect QAOA to have a sub-quadratic speedup in estimating $\mathcal{P}$ for arbitrary graphs, if there exists a quick variational search routine to find $\alpha_j^{\rm opt}$ and $\beta_j^{\rm opt}$. We only plot the total computational time for regular grid graphs in Fig.~\ref{fig: main results}(b-d), because the exponential scaling of the total time with $|E|$ is clean for this class of graphs.

%% file: summary.tex
\section{Summary and Conclusions}\label{sec: summary}

We presented modified AQO and QAOA algorithms to estimate the weighted count of the ground states of an arbitrary classical Hamiltonian, weighted by an arbitrary function. We demonstrated these algorithms using Hamiltonians whose ground states encode edge covers on graphs. We analyzed the computational time required by these algorithms to prepare a quantum system in the ground state of these Hamiltonians, analytically for AQO and numerically for QAOA. We described a statistical technique to estimate the total weight of the ground states, by repeated iterations of AQO or QAOA. We predicted and calculated the scaling properties of the total time taken by these algorithms, and compared this total time against OMCS, which is one of the best error-tractable classical algorithms. We showed that AQO with a linear schedule does not have a speedup over classical OMCS, and that QAOA can have a sub-quadratic speedup over OMCS when the total weight on the ground states is small. We also discussed, with examples, how to minimize the resources required for the variational search of the QAOA parameters, which is crucial for observing the sub-quadratic speedup.

Our ideas solve a long-standing open challenge in quantum optimization of how to count or sample ground states of a classical Hamiltonian with a pre-determined probability distribution. Although we demonstrated our algorithms with counting edge covers, we expect that there are several other problems where our algorithms can provide a competitive advantage over classical algorithms. Several combinatorial counting problems, which have important practical applications such as quantifying and verifying complex systems' performance and uncertainty~\cite{zio2009reliability}, can be cast as ground-state counting problems of Ising-like spin Hamiltonians~\cite{lucas2014ising}. Our work opens avenues to using quantum algorithms to approximately solve such counting problems, even those in the $\#P$-hard complexity class which cannot be approached with existing quantum algorithms for counting~\cite{brassard1998quantum,brassard2002quantum,hen2014fast,wie2019simpler}. Moreover, the ideas we presented have the potential to be implemented on current NISQ devices, and opens avenues to achieving quantum advantage for solving important practical problems in engineering.

%% file: appendix.tex
\section*{Appendix}
\appendix
\section{Proof of Eq.~(\ref{eqn: means})} \label{sec: QMRM}
Here, we derive expressions for $\mean{Q_M}$ and $\mean{R_M}$.

Conditioned on a measurement yielding a ground state, the probability of measuring $\ket{g}$ is $w(g)/\mathcal{P}$.  Then the average weight of one measurement is
\begin{equation}
\mean{R_1} = \sum_{g\in\mathcal{G}} w(g)\frac{w(g)}{\mathcal{P}} = \frac{\mathcal{P}_2}{\mathcal{P}}.
\end{equation}
$\mean{R_M}$ is the average total weight after $M$ measurements. Since each measurement is independent,
\begin{equation}
\mean{R_M} = M\mean{R_1},
\end{equation}
proving the second line of Eq.~(\ref{eqn: means}).

The average number of distinct ground states measured is
\begin{equation}\label{eqn: Qmean}
\mean{Q_M} = \sum_{Q=1}^M Q\ {\rm Pr}(Q_M=Q),
\end{equation}
where ${\rm Pr}(Q_M=Q)$ is the probability of measuring $Q$ distinct ground states in $M$ measurements. To calculate this probability, imagine a set of $M$ experiments where the ground state $\ket{g_1}$ is measured $n_1$ times, $\ket{g_2}$ is measured $n_2$ times, and so on,  such that $n_1+n_2+\cdots + n_Q = M$ and $n_1,\cdots,n_Q\geq1$. The probability that this set of measurements occurs is
\begin{equation}
{\rm Pr}(\{g_i,n_i\}) = \left(\frac{w(g_1)}{\mathcal{P}}\right)^{n_1} \left(\frac{w(g_2)}{\mathcal{P}}\right)^{n_2}\cdots \left(\frac{w(g_Q)}{\mathcal{P}}\right)^{n_Q} \frac{M!}{n_1!n_2!\cdots n_Q!}.
\end{equation}
Then,
\begin{equation}
{\rm Pr}(Q_M=Q) = \sum_{ n_1+n_2+\cdots+n_Q=M} {\rm Pr}(\{g_i,n_i\}).
\end{equation}

We will show that Eq.~(\ref{eqn: Qmean}) leads to the first line of Eq.~(\ref{eqn: means}) in the main text, by comparing the coefficient of the product $ \left(\frac{w(g_1)}{\mathcal{P}}\right)^{n_1} \left(\frac{w(g_2)}{\mathcal{P}}\right)^{n_2}\cdots \left(\frac{w(g_Q)}{\mathcal{P}}\right)^{n_Q} $ in both equations. This coefficient in Eq.~(\ref{eqn: Qmean}) is
\begin{equation}
C_1(n_1,n_2,\cdots n_Q) = Q \frac{M!}{n_1!n_2!\cdots n_Q!}.
\end{equation}
The coefficient of the same product in $\mean{Q_M}$ in Eq.~(\ref{eqn: means}) is
\begin{eqnarray}
C_2&(n_1,n_2,\cdots n_Q) = \sum_{\mu=1}^M (-1)^{\mu-1} \left(\begin{array}{c}M\\ \mu\end{array}\right) \left( \frac{(M-\mu)!}{(n_1-\mu)!n_2!n_3!\cdots n_Q!} \right.\nonumber\\
& +\left. \frac{(M-\mu)!}{n_1!(n_2-\mu)!n_3!\cdots n_Q!} + \cdots \right)\nonumber\\
&= \frac{M!}{n_1!n_2!\cdots n_Q!} \sum_{\mu=1}^M (-1)^{\mu-1} \left( \left(\begin{array}{c}n_1\\ \mu\end{array}\right) + \left(\begin{array}{c}n_2\\ \mu\end{array}\right) + \cdots + \left(\begin{array}{c}n_Q\\ \mu\end{array}\right) \right)\nonumber\\
&= Q\frac{M!}{n_1!n_2!\cdots n_Q!}.
\end{eqnarray}
Therefore, $C_1(n_1,n_2,\cdots n_Q) = C_2(n_1,n_2,\cdots n_Q)$. This proves the first line of Eq.~(\ref{eqn: means}).

Although we have completed the proof for Eq.~(\ref{eqn: means}), we present another, simpler, proof for the first line of Eq.~(\ref{eqn: means}), for the special case $q=1/2$. For this special case,  $w(\phi)=1/2^{|E|}\ \forall \ket{\phi}$ and $\mathcal{P}_\mu = N_0^{(0)}/2^{\mu |E|}$. If the number of distinct ground states measured in $M$ measurements is $Q_M$, then the conditional average number of distinct ground states after one more measurement is 
$\mean{Q_{M+1}} = Q_M + (1-Q_M/\mathcal{P}_0)$.
Then, averaging over all possible values of $Q_M$, we get 
$\mean{Q_{M+1}} = 1 + \mean{Q_M}(1-1/\mathcal{P}_0)$.
This recursive relation leads to a geometric series for $\mean{Q_M}$, whose result is
\begin{equation}
\mean{Q_M} = \sum_{\mu=0}^{M-1} (1-1/\mathcal{P}_0)^\mu  = \mathcal{P}_0 (1-(1-1/\mathcal{P}_0)^M).
\end{equation}
Binomially expanding this equation leads to the first line of Eq.~(\ref{eqn: means}).

\section{Proof of Eq.~(\ref{eqn: delta})} \label{sec: confidence}
Here, we calculate the probability $1-\delta$ that $\mathcal{P}_{\rm est}$ has a maximum relative error $\epsilon$.
\begin{equation}
1-\delta \equiv {\rm Pr}\left(\left|1-\frac{\mathcal{P}_{\rm est}}{\mathcal{P}}\right| < \epsilon\right) = {\rm Pr}\left( \frac{\mean{Q_M}-M\epsilon}{1-\epsilon} < \overline{Q}_M < \frac{\mean{Q_M}+M\epsilon}{1+\epsilon} \right).
\end{equation}
For large enough sample size $S$, the sample mean $\overline{Q}_M$ is normally distributed with $\mean{Q_M}$ and variance ${\rm var}(Q_M)/S$ (due to the central limit theorem). Therefore,
\begin{equation}\label{eqn: central limit theorem}
1-\delta = \frac{1}{2}{\rm erf}\left( \frac{(M-\mean{Q_M})\epsilon}{1+\epsilon}\sqrt{\frac{S}{{\rm var}(Q_M)}} \right) + \frac{1}{2}{\rm erf}\left( \frac{(M-\mean{Q_M})\epsilon}{1-\epsilon}\sqrt{\frac{S}{{\rm var}(Q_M)}} \right).
\end{equation}
The variance of $Q_M$ can be calculated using the same techniques as~\ref{sec: QMRM}, yielding
\begin{eqnarray}\label{eqn: var}
{\rm var} (Q_M) &= M^2 + \sum_{\mu=2}^M (-1)^{\mu-1}(2M-1) \left(\begin{array}{c}M\\ \mu\end{array}\right) \frac{\mathcal{P}_\mu} {\mathcal{P}^\mu} - \mean{Q_M}^2\nonumber\\
&\simeq \frac{M(M-1)}{2} \frac{\mathcal{P}_2}{\mathcal{P}^2} + O(\mathcal{P}_3/\mathcal{P}^3)
\end{eqnarray}
Plugging $\langle Q_M\rangle$ and ${\rm var}(Q_M)$ from Eqs.~(\ref{eqn: means}) and~(\ref{eqn: var}) into Eq.~(\ref{eqn: central limit theorem}) leads to Eq.~(\ref{eqn: delta}).

%% file: algo.tex
\section{Full algorithm for AQO}\label{sec: full algo}
Here, we describe all the subroutines to implement AQO: Determining $T_{\rm AQO}$ in Algorithm~\ref{algo: sub1}, measuring a ground state in one iteration of AQO in Algorithm~\ref{algo: sub2}, and estimating $\mathcal{P}$ in Algorithm~\ref{algo: sub3}.

\begin{algorithm}
\begin{algorithmic}[1]
\STATE Arbitrarily make a guess for $T_{\rm AQO}$.
\STATE Initialize system in $\ket{\psi(0)}$ using the appropriate circuit, such as the one in Fig.~\ref{fig: qtm cct}(a).
\STATE Apply unitary operations $\prod_{j=1}^{T_{\rm AQO}}\ \exp(-i\alpha(t_j)\hH_x \rmd t)\exp(-i\beta(t_j)\hH_z \rmd t)$ to $\ket{\psi(0)}$.
\STATE Measure the system in the computational basis.
\STATE Repeat steps 2-4 to compute $\expectation{\psi(T_{\rm AQO})} {\hat{P}_\mathcal{G}}$ or other implementable metric.
\IF {$\expectation{\psi(T_{\rm AQO})} {\hat{P}_\mathcal{G}} < 1-\eta^2$}
\STATE $T_{\rm AQO} \leftarrow 2*T_{\rm AQO}$
\STATE Go to step 2.
\ENDIF
\caption{Subroutine for determining $T_{\rm AQO}$}\label{algo: sub1}
\end{algorithmic}
\end{algorithm}

\begin{algorithm}
\begin{algorithmic}[1]
\STATE Initialize system in $\ket{\psi(0)}$.
\STATE Set $T_{\rm AQO}$ as determined by Algorithm 1.
\STATE Apply unitary operations $\prod_{j=1}^{T_{\rm AQO}}\ \exp(-i\alpha(t_j)\hH_x \rmd t)\exp(-i\beta(t_j)\hH_z \rmd t)$ to $\ket{\psi(0)}$.
\STATE Measure the system in the computational basis.
\IF {Measurement is not a ground state}
\STATE Discard the measurement.
\STATE Go to step 1.
\ELSE
\STATE Record the measurement.
\ENDIF
\caption{Subroutine for measuring a ground state.} \label{algo: sub2}
\end{algorithmic}
\end{algorithm}

\begin{algorithm}[H]
\begin{algorithmic}[1]
\STATE Arbitrarily pick $S\sim O(1)$ and $M$.
\STATE Repeat Algorithm 2 until $M$ ground states are recorded. Compute $Q_M$ and $R_M$.
\IF {$Q_M=M$}
\STATE $M\leftarrow2M$
\STATE Go to Step 3
\ENDIF
\STATE Estimate $\overline{Q}_M$ and $\overline{R}_M$. Estimate $\mathcal{P}_{\rm est}$ using Eq.~(\ref{eqn: formula for P}), and $\delta$ using Eq.~(\ref{eqn: delta}).
\IF {$1-\delta < $ desired confidence}
\STATE $S\leftarrow2S$
\STATE Go to Step 3.
\ENDIF
\caption{Subroutine for estimating $\mathcal{P}$ from measurements.} \label{algo: sub3}
\end{algorithmic}
\end{algorithm}